%% file: Paper_I.tex
\documentclass[useAMS,usenatbib]{mn2e}

\usepackage[a4paper,margin=2cm]{geometry}

\usepackage[pdftex]{graphicx}
\usepackage{color}
\usepackage{url}
\usepackage{aas_macros}
\usepackage{lastpage}
\usepackage{comment}
\usepackage{amsmath}
\usepackage{booktabs}

\usepackage{type1cm}
\usepackage{eso-pic}
\usepackage{dblfloatfix}

\input{commands}

\title[Merger tree convergence and halo merger rates]{Convergence properties of halo merger trees;\\
halo and substructure merger rates across cosmic history}

\author[Poole et al.]{\parbox[t]{\textwidth}{Gregory B.\ Poole$^{1}$\footnotemark, 
Simon J. Mutch$^{1}$, Darren J. Croton$^{2}$ and Stuart Wyithe$^1$} 
\\ \\
$^1$ School of Physics, University of Melbourne, Parksville, VIC 3010, Australia \\
$^2$ Centre for Astrophysics \& Supercomputing, Swinburne University of Technology, P.O. Box 218, Hawthorn, VIC 3122, Australia 
}

\date{draft version \today}
\pagerange{\pageref{firstpage}--\pageref{lastpage}} \pubyear{2017}

\begin{document}

\label{firstpage}

\maketitle

\begin{abstract}
\input{abstract}
\end{abstract}


\section{Introduction}\label{sec-intro}
\renewcommand{\thefootnote}{\fnsymbol{footnote}}
\setcounter{footnote}{1}
\footnotetext{E-mail: gbpoole@gmail.com}
\input{introduction}

\section{Simulations}\label{sec-simulations}
\input{simulations}

\section{Method}\label{sec-method}
\input{method}

\section{Analysis}\label{sec-analysis}
\input{analysis}

\section{Discussion}\label{sec-discussion}
\input{discussion}

\section{Summary}\label{sec-summary}
\input{summary}

\section*{Acknowledgements}
\input{thanks}

\bibliographystyle{mn2e}
\bibliography{biblio}


\label{lastpage}
\end{document}

%% file: commands.tex
\def\lesssim{\mathrel{\hbox{\rlap{\hbox{\lower4pt\hbox{$\sim$}}}\hbox{$<$}}}}
\def\gtrsim{\mathrel{\hbox{\rlap{\hbox{\lower4pt\hbox{$\sim$}}}\hbox{$>$}}}}
\newcommand{\ltaraw}{$\; \buildrel < \over \sim \;$}
\newcommand{\lta}{\lower.5ex\hbox{\ltaraw}}
\newcommand{\gtaraw}{$\; \buildrel > \over \sim \;$}
\newcommand{\gta}{\lower.5ex\hbox{\gtaraw}}
\def\lsim{\mathrel{\rlap{\lower4pt\hbox{\hskip1pt$\sim$}}
    \raise1pt\hbox{$<$}}}                

\newcommand{\DIFdelFL}[1]{}
\newcommand{\DIFaddFL}[1]{}

\newcommand{\quotes}[1]{``#1''}

\newcommand{\ie}{{\rm i.e.}}
\newcommand{\eg}{{\rm e.g.}}
\newcommand{\etc}{{\it etc.}}
\newcommand{\etal}{{\rm et~al.}}
\newcommand{\cf}{{\it c.f.}}

\newcommand{\LCDM}{$\Lambda$CDM}
\newcommand{\Vmax}{$V_{\rm max}$}

\def\apjl{ApJ}%
\def\apjs{ApJS}%
\def\ao{Appl.~Opt.}%
\def\aap{A\&A}%
\newcommand{\TWOLPTIC}{\textsc{2LPTic}}
\newcommand{\PaNICs}{\textsc{PaNICs}}
\newcommand{\CAMB}{\textsc{CAMB}}
\newcommand{\GADGETtwo}{\textsc{GADGET-2}}
\newcommand{\GADGET}{\textsc{GADGET}}
\newcommand{\Subfind}{\textsc{Subfind}}
\newcommand{\Velociraptor}{\textsc{Velociraptor}}
\newcommand{\ROCKSTAR}{\textsc{ROCKSTAR}}
\newcommand{\Meraxes}{\textsc{Meraxes}}
\newcommand{\gbpTrees}{\textsc{gbpTrees}}

\newcommand{\Millennium}{\textit{Millennium}}
\newcommand{\MillenniumII}{\textit{Millennium{-}II}}
\newcommand{\Tiamat}{\textit{Tiamat}}

\newcommand{\TinyTiamat}{\textit{Tiny Tiamat}}
\newcommand{\MediTiamat}{\textit{Medi Tiamat}}
\newcommand{\Tiamatcontrol}{\textit{Tiamat-125}}
\newcommand{\TiamatLR}{\textit{Tiamat-125-LR}}
\newcommand{\TiamatNR}{\textit{Tiamat-125-NR}}
\newcommand{\TiamatMR}{\textit{Tiamat-125-MR}}
\newcommand{\TiamatHR}{\textit{Tiamat-125-HR}}
\newcommand{\GiggleZ}{\textit{GiggleZ}}
\newcommand{\GiggleZLR}{\textit{GiggleZ-LR}}
\newcommand{\GiggleZNR}{\textit{GiggleZ-NR}}
\newcommand{\GiggleZMR}{\textit{GiggleZ-MR}}
\newcommand{\GiggleZHR}{\textit{GiggleZ-HR}}
\newcommand{\GiggleZcontrolall}{\textit{GiggleZ-LR/NR/MR/HR}}
\newcommand{\GiggleZcontrol}{\textit{GiggleZ-control}}
\newcommand{\GiggleZmain}{\textit{GiggleZ-main}}

\newcommand{\eqx}[1]{{${=}{#1}$}}
\newcommand{\simx}[1]{{${\sim}{#1}$}}

\newcommand{\lsimx}[1]{{${\lesssim}{#1}$}}
\newcommand{\gtrx}[1]{{${>}{#1}$}}

\newcommand{\gsimx}[1]{{${\gtrsim}{#1}$}}
\newcommand{\mone}{{${-}{1}$}}
\newcommand{\rmsub}[2]{{$#1_{\rm #2}$}}
\newcommand{\TTTP}[1]{{${10}^{#1}$}}
\newcommand{\scinote}[2]{${#1}{\times}{{10}^{#2}}$}

\newcommand{\tdyn}{$t_{\rm dyn}$}
\newcommand{\tHubble}{$t_{\rm Hubble}$}

\newcommand{\Deltatau}{{${\Delta}{\tau}$}}
\newcommand{\Deltatausnap}{\rmsub{\Delta\tau}{snap}}
\newcommand{\Deltatausearch}{\rmsub{\Delta\tau}{search}}
\newcommand{\zx}[1]{{${z}_{\rm{#1}}$}}
\newcommand{\ax}[1]{{${a}_{\rm{#1}}$}}
\newcommand{\zxeqx}[2]{{\zx{#1}}{\eqx{#2}}}
\newcommand{\zxltx}[2]{{\zx{#1}}{$<$}{#2}}
\newcommand{\zxgtrx}[2]{{\zx{#1}}{\gtrx{#2}}}

\newcommand{\haloid}[3]{{#1$#2$#3}}
\newcommand{\halochainijk}[2]{\haloid{#1}{#2}{}{$\rightarrow$}\haloid{#1}{#2}{'}{$\rightarrow$}\haloid{#1}{#2}{"}}

\newcommand{\halochainik}[2]{\haloid{#1}{#2}{}{$\rightarrow$}\haloid{#1}{#2}{"}}
\newcommand{\halochain}[6]{\haloid{#1}{#2}{#3}{$\rightarrow$}\haloid{#4}{#5}{#6}}

\newcommand{\Deltafgood}{{$\Delta f_{\rm g}$}}
\newcommand{\fmoment}[1]{$f^{{(}#1{)}}_{\rm{g}}$}
\newcommand{\Smoment}[2][]{$S^{{(}#2{)}}_{\rm{#1}}$}
\newcommand{\Smomentmax}[1]{$S^{{(}#1{)}}_{\rm{max}}$}
\newcommand{\nrank}{$m$}
\newcommand{\nsearch}{{$n_{\rm search}$}}
\newcommand{\nsnap}{{$n_{\rm snap}$}}
\newcommand{\nparticle}{{$n_{\rm p}$}}

\newcommand{\oneway}{{$1$-way}}
\newcommand{\twoway}{{$2$-way}}

\newcommand{\Mpc}{\mbox{Mpc}}

\newcommand{\Gpchunit}{$h^{-1} {\rm Gpc}$}
\newcommand{\Mpchunit}{$h^{-1} {\rm Mpc}$}

\newcommand{\kpchunit}{$h^{-1} {\rm kpc}$}

\newcommand{\Msol}{${{\rm M}_{\sun}}$}
\newcommand{\Msolhunit}{${{h^{-1}}{\rm M}_{\sun}}$}

\newcommand{\Gyrs}{\mbox{Gyrs}}


%% file: abstract.tex
We introduce \gbpTrees: an algorithm for constructing merger trees from cosmological simulations, designed to identify and correct for pathological cases introduced by errors or ambiguities in the halo finding process.  \gbpTrees\ is built upon a halo matching method utilising pseudo-radial moments constructed from radially-sorted particle ID lists (no other information is required) and a scheme for classifying merger tree pathologies from networks of matches made to-and-from haloes across snapshots ranging forward-and-backward in time.  Focusing on \Subfind\ catalogs for this work, a sweep of parameters influencing our merger tree construction yields the optimal snapshot cadence and scanning range required for converged results.  Pathologies proliferate when snapshots are spaced by \lsimx{0.128} dynamical times; conveniently similar to that needed for convergence of semi-analytical modelling, as established by Benson \etal\  Total merger counts are converged at the level of \simx{5}\% for friends-of-friends (FoF) haloes of size \nparticle\gsimx{75} across a factor of 512 in mass resolution, but substructure rates converge more slowly with mass resolution, reaching convergence of \simx{10}\% for \nparticle\gsimx{100} and particle mass $m_{\rm p}{\lsim}10^{9}$\Msol.  We present analytic fits to FoF and substructure merger rates across nearly all observed galactic history ($z{\le}8.5$).  While we find good agreement with the results presented by Fakhouri \etal\ for FoF haloes,  a slightly flatter dependance on merger ratio and increased major merger rates are found, reducing previously reported discrepancies with extended Press-Schechter estimates.  When appropriately defined, substructure merger rates show a similar mass ratio dependance as FoF rates, but with stronger mass and redshift dependencies for their normalisation.

%% file: introduction.tex
Large spectroscopic galaxy surveys are a tool of fundamental importance to studies in cosmology and galaxy formation.  With the great strides that have been made in recent years towards surveys of larger volume, depth and redshift coverage, our measurements of fundamental cosmological properties of the Universe -- such as its evolving expansion rate, mass-energy composition and matter power spectrum -- have experienced continual improvements in statistical precision and steady reductions in systematic uncertainties.  In parallel to these advances, tremendous progress in our understanding of galaxy formation has also been made on the back of large surveys.  

The next generation of surveys anticipated to come from DES, LSST, and Euclid may face an entirely new obstacle potentially inhibiting future progress: systematic uncertainties in the theoretical models required to interpret or even extract information.  Indeed, current estimates for Euclid are that simulations of mock observed matter power spectra covering scales from $20$ \kpchunit\ to $1000$ \Mpchunit\ with both statistical \emph{and} systematic uncertainties of less than one percent will be required to fully realise the potential of the mission for several key questions in cosmology \citep[see][for a discussion]{Schneider:2016p2667}.  Control of systematic uncertainties will require, amongst other things, mock catalogs covering enormous volumes with sufficient fidelity to account for selection effects at this level of accuracy.

Two principle numerical tools have historically found application in extragalactic modeling: hydrodynamic cosmological simulations and semi-analytic galaxy formation models (SAMs) applied to halo catalogs and merger trees built from collisionless cosmological simulations.  Both have played important roles in building our current understanding of galaxy formation with a principle asset of SAMs being the large volumes they can be applied to.  As such, they are likely to continue playing an important role in the coming era of massive surveys.

There are several steps involved in building a mock galaxy catalog with a semi-analytic model.  The first and most robustly studied\footnote{However, note that despite the tremendous attention that has been given over recent decades to ensure that efficient algorithms for generating enormous collisionless N-body simulations remain precise, baryonic processes which inhibit halo growth are neglected by these algorithms.  To achieve the maximal levels of precision attainable by semi-analytic modelling, such processes will need to be taken into account.  See \citet{2013MNRAS.431.1366S,2015MNRAS.451.1247S,2017MNRAS.467.1678Q}, for more details.} is the construction of a large collisionless cosmological N-body simulation.  The result of such a simulation is a series of \quotes{snapshots} depicting the instantaneous positions, velocities and identities of all simulation particles.  A halo finding procedure is then applied which seeks to identify and report relevant physical properties (\eg\ masses, angular momenta, velocity dispersions, \etc) for all distinct and bound structures representing the presumptive sites of galaxy formation.  Finally, the identities of these bound structures need to be connected from one snapshot to the next in the form of merger trees, yielding full physical accounts of the mass accretion histories of galaxy formation sites.  Merger trees and halo properties form the fundamental input to semi-analytic galaxy formation models.

Unfortunately, there are many tools and approaches available for each and every one of these steps, resulting in an enormous proliferation of models.  While there is broad agreement in some regards amongst these models, many important details still demand further study, with supposedly similar models yielding significant differences.  These differences can even lead to qualitatively different conclusions about the influence of physical processes. 

Jointly, these matters have recently spurred a series of studies seeking to establish, quantify and understand the systematic biases introduced by each of these stages in building a SAM \citep[see papers from the \quotes{Haloes gone MAD} and related projects:][for example]{Knebe:2011p1277,Onions:2012p2662,Onions:2013p2663,Elahi:2013p2672,Pujol:2014p2673,Hoffmann:2014p2674,Behroozi:2015p2660}.  Of all the steps involved in deploying semi-analytic models on collisionless N-body simulations, the one that has received the least amount of attention is the construction of merger trees.  While conceptually simple, errors, ambiguities and systematics in the halo finding process can introduce pathological errors, making the construction of merger trees a significant technical challenge.  A plethora of merger tree construction methods and implementations have been presented in the literature \cite[\eg][see \citealt{Srisawat:2013p2512} for a brief discussion and comparison of these and other methods]{Jung:2014p2678,Jiang:2014p2677,Behroozi:2013p2471,Han:2012p2515,Springel:2005a}.  A recent series of papers (a product of the \quotes{Sussing Merger Trees} comparison project) have taken some initial steps towards revealing: the extent to which these approaches agree \citep{Srisawat:2013p2512}; their dependance on different methods of halo identification \citep{Avila:2014p2675}; and the cascading consequences for semi-analytic galaxy formation models \citep{Wang:2016p2676}.  Limited to some degree by only considering comparisons that are possible across all of the very different approaches studied, a great deal of systematic study remains to be done regarding the detailed nature of merger tree pathologies (a discussion that is difficult to conduct when comparing methods that handle such issues in fundamentally different ways).  Furthermore, additional study regarding convergence properties with (for example) mass resolution and snapshot cadence is warranted.  The latter of these is of particular concern for studies at high redshift, where simulation programs often have sparse snapshot coverage during an epoch when dynamical times are short and merger rates very high \citep[\eg][]{Poole:2016p2664}.

With this paper we introduce the \gbpTrees\ merger tree algorithm and present a rigorous test of its convergence properties when applied to catalogs constructed with the popular \Subfind\ halo finding code \citep{Springel:2001}.  \gbpTrees\ explicitly isolates and labels pathological situations in halo trees permitting detailed population studies, including their convergence with (for example) the method's tuneable parameters and with simulation snapshot cadence and mass resolution.  With ideal values for the parameters controlling the performance of our merger tree algorithm determined, we then deploy it on a series of simulations spanning the range of masses most relevant to galactic studies to determine the mass-ratio, mass and redshift dependance of merger rates for both friends-of-friends (FoF) haloes and substructures (as identified by \Subfind) across cosmic time.

In Section \ref{sec-simulations} we present the suites of simulations that will be used to perform our convergence study and to subsequently quantify halo merger rates.  In Section \ref{sec-method} we present a detailed motivation for the design and details of the implementation of our merger tree algorithm.  We subsequently explore the convergence properties of the algorithm (when applied to \Subfind\ catalogs) in Section \ref{sec-analysis} and quantify merger rates for $z{\le}8.5$ in a Planck-2015 cosmology.  We conclude with a brief discussion in Section \ref{sec-discussion} and provide a summary of our findings in Section \ref{sec-summary}.

Throughout this paper we will employ simulations run with two similar cosmologies.  Discussions involving the design and implementation of our merger tree code and analyses of convergence behaviour in Sections \ref{sec-method} and \ref{sec-analysis} will employ a standard spatially-flat WMAP-5 cosmology \citep{Komatsu:2009} ($h$, $\Omega_{\rm{m}}$, $\Omega_{\rm{b}}$, $\Omega_\Lambda$, $\sigma_8$, $n_{\rm{s}}$)${=}$($0.727$, $0.273$, $0.0456$, $0.705$, $0.812$, $0.96$) while the merger rates presented in Section \ref{sec-analysis} are extracted from simulations run with a standard spatially-flat Planck $\Lambda$CDM cosmology based on 2015 data \citep[][]{PlanckCollaboration:2015p2562} ($h$, $\Omega_{\rm{m}}$, $\Omega_{\rm{b}}$, $\Omega_\Lambda$, $\sigma_8$, $n_{\rm{s}}$)${=}$($0.678$, $0.308$, $0.0484$, $0.692$, $0.815$, $0.968$).

Throughout this work, intervals in time will be expressed in units of a halo's dynamical time at its virial radius.  This it taken to be 10\% of the Hubble time (\tHubble${=}1/H(z)$, with $H(z)$ being the Hubble expansion rate) at any given redshift\footnote{Halo virial radii are generally defined in terms of a mean density $\bar{\rho}{=}\Delta\rho_c$ (with $\Delta{\sim}200$ being the virial overdensity parameter and $\rho_c{=}3{H(z)}^2/8\pi G$ the critical density at redshift $z$, with $G$ being Newton's gravitational constant).  At this radius ($R_{200}$), the characteristic orbital speed is ${v_{200}}^2{=}GM_{200}/R_{200}$ (with $M_{200}{=}4\pi {R_{200}}^3\bar{\rho}/3$ being the mass within the virial radius) and crossing times can be approximated as $t_{\rm cross}{\sim}R_{200}/v_{200}$.  Taking this measure of crossing time as the dynamical time and combining these relations yields \tdyn${=}0.1/H(z)$.}.  Times in this dimensionless system of units will be denoted by $\tau$.  At all redshifts, the Hubble time is $\tau{=}10$ in this system.

\begin{table*}
\begin{minipage}{170mm}
\begin{center}
\begin{tabular}{ccccccc}
\hline
Simulation      &
$L$ [\Mpchunit]     & 
$N_{\rm p}$   & 
$m_{\rm p}$  [$h^{-1}$M$_\odot$] &
$\epsilon$ [\kpchunit] &
snapshots  &
cosmology \\
\hline
\GiggleZLR	&	125	&	135$^3$	&	\scinote{6.01}{10}	&	18.5	&	931 to \zx{}\eqx{0}	&	WMAP-5\\
\GiggleZNR	&	125	&	270$^3$	&	\scinote{7.52}{9}	&	9.3	&	931 to \zx{}\eqx{0}	&	WMAP-5\\
\GiggleZMR	&	125	&	540$^3$	&	\scinote{9.40}{8}	&	4.6	&	466 to \zx{}\eqx{0}	&	WMAP-5\\
\GiggleZHR	&	125	&	1080$^3$	&	\scinote{1.17}{8}	&	2.3	&	234 to \zx{}\eqx{0}	&	WMAP-5\\
\hline                                                                             
\TiamatLR		&	125	&   135$^3$	&	\scinote{6.79}{10}	&	18.5	&	250 to \zx{}\eqx{0}	&	Planck-2015\\
\TiamatNR	&	125	&   270$^3$	&	\scinote{8.48}{9}	&	9.3	&	250 to \zx{}\eqx{0}	&	Planck-2015\\
\TiamatMR	&	125	&   540$^3$	&	\scinote{1.06}{9}	&	4.6	&	250 to \zx{}\eqx{0}	&	Planck-2015\\
\TiamatHR	&	125	& 1080$^3$	&	\scinote{1.33}{8}	&	2.3	&	250 to \zx{}\eqx{0}	&	Planck-2015\\
\hline                                                                             
\Tiamat		&	67.8	&   2160$^3$	&	\scinote{2.64}{6}	&	0.628	&	164 to \zx{}\eqx{1.8}	&	Planck-2015\\
\MediTiamat	&	25	&   1080$^3$	&	\scinote{7.83}{5}	&	0.463	&	101 to \zx{}\eqx{5}	&	Planck-2015\\
\TinyTiamat	&	10	&   1080$^3$	&	\scinote{6.79}{4}	&	0.185	&	101 to \zx{}\eqx{5}	&	Planck-2015\\
\hline
\end{tabular}
\caption{Box sizes ($L$), particle counts ($N_{\rm p}$), particle mass ($m_{\rm p}$), gravitational softenings ($\epsilon$), snapshot coverage and cosmologies for the simulations used in this work. \label{table-simulation_parameters}}
\end{center}
\end{minipage}
\end{table*}

%% file: simulations.tex
For this paper we will employ three simulation suites: the \GiggleZcontrol\ simulations previously presented in \citet{Poole:2015a}, the \Tiamat\ simulations previously presented in \citet[][including an extension of the main \Tiamat\ simulation to \zxeqx{}{1.8}]{Poole:2016p2664} and a new set of simulations (which we shall refer to as the \Tiamatcontrol\ simulations) run with the same updated Planck-2015 cosmology and snapshot strategy as the \Tiamat\ simulations but with identical volumes, resolutions and the same matched-phase scheme employed by the \GiggleZcontrol\ simulations.

The \GiggleZcontrol\ simulations will be used for most of our convergence studies due to their large numbers of snapshots, allowing merger trees with very fine temporal resolution to be constructed.  The \Tiamat\ simulations provide an excellent resource for building highly accurate trees at redshifts \zxgtrx{}{5} over the full range of halo masses relevant to galaxy formation during the epoch of reionization.  Our new \Tiamatcontrol\ simulations are an effective low-redshift companion to previously published \Tiamat\ simulations, since they possesses the same cosmology and snapshot cadence but with a mass resolution (in the \TiamatHR\ case at least) and volume adequate for describing a wide range of halo masses relevant to galaxy formation at low redshift.  Furthermore, as with previous \Tiamat\ simulations, we have access to the full substructure hierarchy for the \Tiamatcontrol\ simulations (which we do not have access to for the \GiggleZ\ simulations) permitting the construction of inclusive substructure masses, needed for comparisons of subhalo populations between simulations run at differing mass resolutions.  This will be needed for explorations of mass-resolution convergence of substructure merging properties, for example.

\subsection{\GiggleZ}

The \GiggleZ\ suite consists of 5 simulations: a large \GiggleZmain\ run consisting of 2160$^3$ particles distributed in a periodic box $1$ \Gpchunit\ on a side, and 4 simulations of a 125 \Mpchunit\ \quotes{control} volume spanning a factor of 512 in mass resolution with mean snapshot temporal resolutions as fine as \simx{0.015} dynamical times.  All were initialised, run and analysed assuming a standard WMAP-5 \LCDM\ cosmology, with parameters taken from \citet{Komatsu:2009}: ({$\Omega_\Lambda$}, {$\Omega_M$}, {$\Omega_b$}, {$h$}, {$\sigma_8$}, {$n$})=(0.727, 0.273, 0.0456, 0.705, 0.812, 0.960).

Initial conditions for all runs were generated using the bespoke code \PaNICs\ (Parallel N-body Initial Conditions) which follows the approach of \citet{Bertschinger:2001} to construct displacement fields and a first-order Zel'dovich approximation \citep{ZelDovich:1970,Buchert:1992} to construct particle velocities.  Displacement fields were constructed to yield a matter power spectrum given by the Boltzmann code \CAMB\ \citep{Lewis:2000} run with the WMAP-5 cosmological parameters given above.

All simulations were evolved from these initial conditions to \zxeqx{}{0} using the publicly available N-body code \GADGETtwo\ \citep{Springel:2005b}; a Tree-Particle Mesh (TreePM) code well suited to large distributed memory systems.  Standard run parameters were used, with gravitational softenings chosen to be $2$\% of the mean inter-particle spacing of each run and a \GADGETtwo\ integration parameter accuracy of $\eta{=}0.01$, as motivated by the convergence study presented in \citet{Poole:2015a}.

The first task in generating merger trees is halo finding: the identification of gravitationally bound substructures within a simulation which we can identify as the sites of galaxy formation.  As with all simulations used in this paper, this has been done for \GiggleZ\ using the configuration-space halo finder \Subfind, which utilises a friends-of-friends (FoF) algorithm to identify coherent overdensities of particles and a configuration-space substructure analysis to determine bound overdensities within each FoF halo.  The linking length parameter used for the FoF halo finding was set to the standard value of $0.2$.  The raw output of this process is a list of particle IDs for each FoF and substructure halo.  Conveniently, \Subfind\ reports these particles in order of binding energy; a fact that we will make good use of when constructing halo matches between the snapshots of our simulations.  The method we describe in this paper is capable of constructing merger trees for both the FoF haloes and substructures of our simulations and of tracking the full substructure hierarchy of each halo.  In what follows, we will speak generically of both sets of structures as \quotes{haloes}, with the understanding that this refers to FoF haloes when building FoF halo trees and substructures when building substructure trees, unless otherwise stated.

It is important to emphasise that the results we report throughout this present work will -- in many cases -- be specific to analyses conducted when using configuration-space halo finders and indeed, may be specific to analyses conducted using the \Subfind\ implementation of this approach.  Results may be very different when applied to catalogs generated from phase-space halo finders, for example \citep[see][for a list and comparison of commonly used codes]{Knebe:2011p1277}.  Our tree construction algorithm is generic however, operating only on particle lists sorted in some radial fashion, and can be applied straight-forwardly to any halo finder that generates such tables.  Future work will examine systematic differences expected from various halo finding approaches.

For this present work, we will focus specifically on the four 125 \Mpchunit\ \quotes{control volume} simulations \GiggleZcontrolall\ for our convergence analyses.  These simulations were constructed with two qualities which make them particularly useful for such work: the phases of the fluctuations imprinted on their initial conditions were constructed to represent an identical volume (thus minimising the influence of cosmic variance in direct comparisons), and a large number of snapshots were written to permit a careful study of the convergence of our merger tree properties through changes in snapshot cadence.

 The basic specifications for these 4 runs are listed in Table \ref{table-simulation_parameters}.  Interested readers should refer to \citet{Poole:2015a} for additional details regarding their construction.

\subsection{\Tiamat}

The \Tiamat\ simulations were run to support the Dark-ages Reionization and Galaxy-formation Observables from Numerical Simulations (DRAGONS) program; a galaxy formation and cosmic reionization simulation campaign featuring \Meraxes, a semi-analytic model which fully and self-consistently couples galaxy formation physics to the ionizing UVB feedback field produced by young galaxies in the early Universe \citep{Mutch:2016p2671}.

The \Tiamat\ Suite consists of three simulations: two smaller cubic periodic volumes (one 10 \Mpchunit\ box and one 22.6 \Mpchunit\ box) represented by 1080$^3$ particles and the main \Tiamat\ run consisting of 2160$^3$ particles in a cubic periodic volume of 100 \Mpc.  These simulations -- like all those utilised in this paper -- were evolved from initial conditions using the \GADGET\ code.  Initial conditions were established at \zxeqx{}{99} using the 2nd-order Lagrangian perturbation code \TWOLPTIC\ \citep[][with fixes to ensure correct behaviour when particle or displacement field grid cell counts exceed $2^{32}{-}1$]{Crocce:2006p2494}.  Particle loads and displacement fields were taken to be regular periodic grids of dimensions 2160$^3$ for \Tiamat\ and 1080$^3$ for all other cases.  The input power spectrum was generated using CAMB \citep{Lewis:2000} with the $\Lambda$CDM parameters appropriate to a Planck-2015 cosmology \citep[][]{PlanckCollaboration:2015p2562} ($h$, $\Omega_{\rm{m}}$, $\Omega_{\rm{b}}$, $\Omega_\Lambda$, $\sigma_8$, $n_{\rm{s}}$)${=}$($0.678$, $0.308$, $0.0484$, $0.692$, $0.815$, $0.968$).  Halo finding was performed with \Subfind.

For this work we have extended the \Tiamat\ simulation from the final time step at \zxeqx{}{5} presented in \citet{Poole:2016p2664} down to \zxeqx{}{1.8}.  Our snapshot strategy was changed for this portion of the run from step sizes constant in proper time to step sizes constant in dynamical time.  This represents an additional 63 snapshots in addition to the original 101. 

The basic specifications for these 3 runs are listed in Table \ref{table-simulation_parameters}.   Interested readers should refer to \citet{Poole:2016p2664} for additional details regarding their construction.

\subsection{\Tiamatcontrol}

The \Tiamatcontrol\ simulations have been constructed to provide a low-redshift compliment to the \Tiamat\ simulations.  Precisely the same approach to constructing initial conditions was taken, using the same cosmology and snapshot redshifts.  The transition to snapshot timesteps constant in fractional dynamical time is continued all the way to \zxeqx{}{0} representing 150 snapshots at \zxltx{}{5} for a total of 251 snapshots.  Box sizes and mass resolutions are chosen to match the \GiggleZcontrol\ simulations (modulo the natural differences due to cosmology changes) with phases again selected to be identical between the simulations, moderating the effects of cosmic variance when trends with mass resolution are explored.

The basic specifications for these 4 runs are listed in Table \ref{table-simulation_parameters}.

%% file: method.tex
With our simulations run and their haloes identified, construction of merger trees proceeds through a process of progressively tracking the unique identities of every halo in consecutive simulation snapshots.  This is achieved through a process of matching which proceeds from the top of each tree (\ie\ its {\it root}), starting at the simulation's last snapshot (generally redshift $z{=}0$), to the {\it leaves} of the tree where haloes first become identified at higher redshifts.  To a first approximation, the underlying assumption throughout this process is that systems can merge but not subdivide, although we will show later that haloes do in fact break apart.  This can be due to physical processes (for instance, through tidal stripping) or can be due to problems in the halo finding process (nearby haloes can be briefly joined, only to break apart afterwards).  We generically refer to these and other abnormal cases as \quotes{tree pathologies} and our method is designed to identify when they happen.

For a simulation with \nsnap\ output snapshots at expansion factors $a_i$ and cosmic times $t_i$ (where $i$ runs from $0$ for the simulation's first and highest-redshift snapshot to \nsnap\mone\ at its last and lowest-redshift snapshot), our trees are constructed in reverse chronological order, identifying the haloes at $a_{{\rm n_{snap}}-1}$ as merger tree roots then matching haloes at $a_{{\rm n_{snap}}-2}$ to haloes at $a_{{\rm n_{snap}}-1}$, etc. We follow the common practice of denoting a halo's successors as {\it descendants} and predecessors as {\it progenitors}.

We point out here that the most computationally expensive step in our merger tree method is the construction of matches between snapshots.  As we will motivate below, this needs to be performed -- not just between consecutive snapshots $i$ and $i{+}1$ -- but between snapshots $i$ and $i{+}$\nsearch\ (with \nsearch\ as large as $128$ used in this work) and not just in one direction but in two for all snapshots $i{=}0$ to \nsnap\mone.  To ensure practical run-times for this part of the calculation -- even for our largest simulations -- this part of the code has been fully parallelised using the Message Passing Interface (MPI).

\subsection{Halo Matching}

The first and most fundamental exercise in our method is the construction of unique \quotes{matches} for each halo in one snapshot to one in another snapshot.  To achieve this, we use only the halo particle ID lists reported by the halo finder, which we take to be radially sorted in some way from the centre outwards (this is done automatically by \Subfind, which reports halo particle ID lists in order of decreasing gravitational binding energy).  No velocity information is used and the only spatial information employed is that implicitly encoded in the radial sorting of the halo particle IDs.  As we shall argue, it is critical to optimise our matching algorithm to follow the core material of each halo.  To achieve this we shall use the (sorted) particle ordering to centrally weight our matches.

\subsubsection{Complicating situations}\label{sec-complicating_situations}

\begin{figure}
\includegraphics[width=80mm]{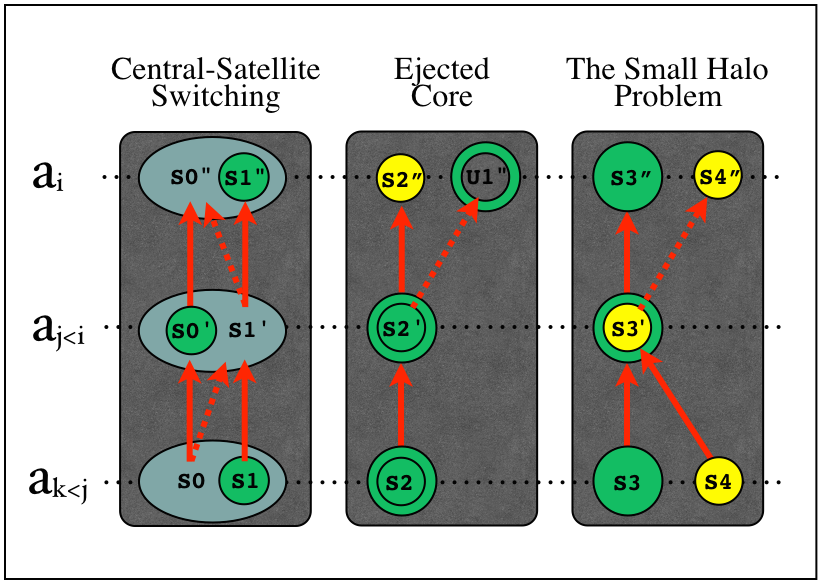}
\caption[Cases which complicate matching]{A figure illustrating three important cases complicating halo-to-halo matching which our merger tree algorithm is designed to address: central-satellite switching, ejected cores and the small halo problem.  Dotted horizontal lines represent progressive simulation snapshots $a_i$ where $i$ runs from $0$ to \nsnap\mone\ in order of increasing expansion factor.  Solid red arrows show the matches which we want to make and red dotted arrows show the incorrect results that may occur when matching and tree construction is not properly optimised.  In the case of \emph{central-satellite switching}, the dense substructure which our halo finder reports as the central halo at one time step (\eg\ \haloid{S}{0}\ at $a_{\rm k}$) becomes a satellite at the next (\haloid{S}{0}{'} at $a_{\rm j}$), at which time another halo becomes central (\haloid{S}{1}{'}).  This can occur several times in succession during merger events and may involve more than two haloes.  In the case of \emph{ejected cores}, the central core of a system can become decoupled from its outer halo (\eg\ \haloid{S}{2}{"} at $a_{\rm i}$).  In this case it is important that the progenitor line follow this ejected core, and not its outer halo (the unwanted pointer is illustrated by the dotted line from \haloid{S}{2}{'}{$\rightarrow$}\haloid{U}{1}{"}).  In the case of the \emph{small halo problem}, a small substructure (in this case, \haloid{S}{4}; perhaps even as small as our halo resolution limit) can pass transiently through the centre of another halo (\haloid{S}{3}{'}; this halo may be many orders of magnitude more massive).  Due to the central weighting of our matching, care must be taken to prevent the progenitor line of \haloid{S}{3}\ from following that of this interloper (the unwanted pointer is illustrated by the dotted line from \haloid{S}{3}{'}{$\rightarrow$}\haloid{S}{4}{"}).\label{fig-matching_complication_cases}} 
\end{figure}

\begin{figure*}
\begin{minipage}{175mm}
\begin{center}
\includegraphics[width=150mm]{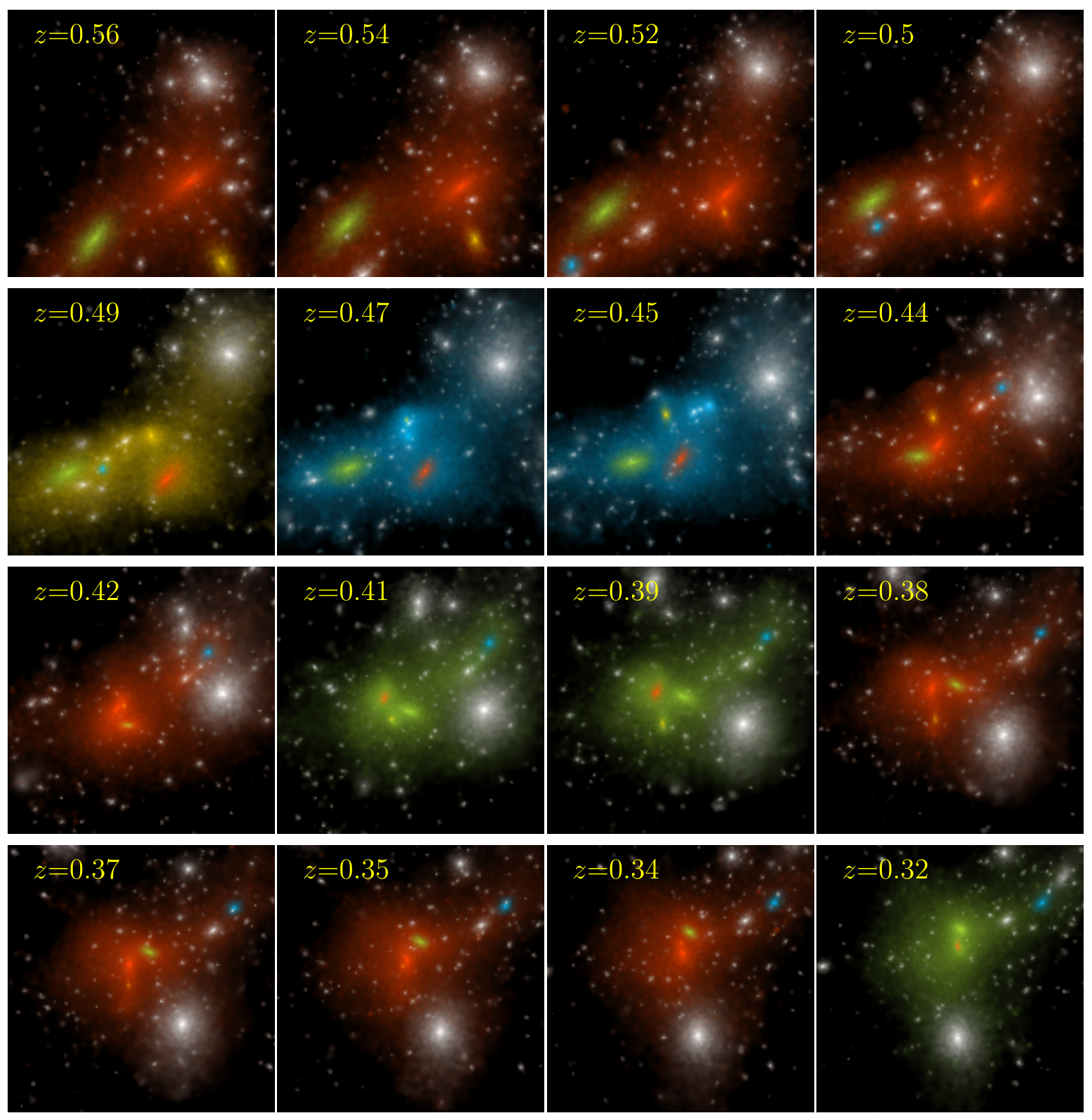}
\end{center}
\caption[Fiducial system time sequence]{A figure depicting our \quotes{fiducial} system -- a many-body interaction in the \GiggleZHR\ simulation -- which we use to illustrate several concepts in this paper.  Colours highlight material from several of the most important systems in the interaction.  Images are $1.25$ \Mpchunit\ on a side with yellow text indicating the redshift of each panel.  Movies depicting this system can be found at the following link: \url{http://www.astronomy.swin.edu.au/~gpoole/gbpTrees.html} \label{fig-fiducial_array}} 
\end{minipage}
\end{figure*}

Because we are primarily interested in science questions involving galactic evolution, our method is optimised to track the cores of haloes where galaxies are most likely to form and evolve.  This is a very important distinction with significant consequences since -- as we will discuss in this subsection -- the cores of haloes do not necessarily follow the majority of their halo material, particularly during complicated dynamical disturbances.  This choice has many fundamentally important consequences for how we build matching identities for haloes across snapshots and for the logic we employ when identifying and treating merger tree pathologies.  

In Figure \ref{fig-matching_complication_cases} we illustrate three particularly important related problems.  These are specific cases of what we shall define below as \quotes{bridged} haloes, which demand particularly careful consideration of (and which ultimately motivate) our halo matching strategy:

\emph{Central-satellite switching}\footnote{Referred to as host-subhalo swaps in some recent studies \citep[][for example]{Behroozi:2015p2660}, we adopt this language here to stay consistent with the nomenclature typically adopted by the \Subfind\ literature.}: During the later stages of a merger, the cores of interacting haloes often continue to orbit each other within a common envelope of material.  By design, \Subfind\ identifies one substructure as a FoF group's \quotes{central} and this system inherits the majority of this common envelope of material.  Other substructures are identified as \quotes{satellites} consisting only of the core of its original merging system.  However, the remnant core identified as the system's central can alternate between two-or-more substructures leading to very large transient fluctuations in their masses.  This can happen many times over the course of an evolving merger.

In Figure \ref{fig-fiducial_array} we present the evolution of a many-body merger between several \simx{10^{12}} to \simx{10^{13.5}} \Msolhunit\ systems which ultimately merge to form a $10^{14}$ \Msolhunit\ remnant (we call this our \quotes{fiducial} system, and will refer to it later as well).  The colour changes of the extended envelope of the system reflects moments where a change has occurred in the satellite identified as the central halo of the system following a central-satellite switching event. 

A matching approach which seeks to connect a halo to a progenitor simply by maximising the amount of shared material will incorrectly swap the identities of systems evolving through such fluctuations.  A main progenitor line for the most massive incoming system will be created which follows the outer envelope of material.  As multiple systems take turns as the central halo, branches will become broken and progenitor lines for those haloes will be undefined.  In later sections we shall develop our matching method which utilises a system of centrally weighted matches to robustly follow halo cores through these events.  


\emph{Ejected cores}: A related problem to the case of central-satellite switching (indeed, central-satellite switching can be thought of as a special case) is the decoupling of a halo's core from its outer material during dynamical disturbances.  In these cases, it is important for our matching algorithm to unambiguously follow this ejected core rather than the decoupled outer halo.

\emph{The small halo problem}: Occasionally (particularly for very well resolved systems) very small substructures can pass directly through the centres of larger systems.  Centrally weighted matches can be incorrectly influenced -- or even dominated -- by the particles of such interloping systems.  This is a particularly serious problem for halo catalogs constructed with configuration-space halo finders such as \Subfind.

To conclude: to minimise the effects of these complicating situations we need an approach to halo matching optimised to robustly follow the bulk of a halo's core material while avoiding being dominated by a large amount of outer-halo material or a small amount of central material.

\subsection{Quantifying halo matches with pseudo-moments}

With the pitfalls of halo matching discussed and the need for a central matching scheme motivated, we now discuss our method for quantifying halo matches.  

Since particles can diffuse in or out of a halo over time or be exchanged during interactions, a halo can sometimes have several prospective matches.  For each halo $\alpha$ with \nparticle\ particles in snapshot $i$ identified in a set of haloes $\beta{=}\{\beta_{\rm k}\}$ in a specific snapshot $j{\ne}i$, we will perform our match selections using pseudo-radial moments computed from the sorted particle lists using the following equation:
\begin{equation}\label{eqn-score}
S^{{(}m{)}}_{\rm{k}}{=}\sum\limits_{l{=}1}^{n_k} {r}^{m}
\end{equation}

\noindent where $r$ is the rank-order (with indexing starting at $1$) of the particle in halo $\alpha$ with the sum running over all the \rmsub{n}{k} particles shared by $\alpha$ and \rmsub{\beta}{k}.  In the trivial case of \nrank\eqx{0}, \Smoment[k]{m} simply becomes \rmsub{n}{k}; the number of particles involved in the prospective match.  A centrally weighted sum can be achieved from Equation \ref{eqn-score} by selecting \nrank\ to be less than zero.  We have experimented with several values of \nrank, finding a value of \nrank\eqx{-1.5} to be too steep to be useful (\Smoment[k]{-1.5} saturates at a value of ${\sim}2.6$ in this case, rendering results highly unstable to cases where even just the $r$\eqx{1} particle is absent from the match, even for very high values of \rmsub{n}{k}) and \nrank\eqx{-\frac{2}{3}} \citep[as used by][in their analysis of the \MillenniumII\ simulation]{BoylanKolchin:2009p2666} to be too shallow (\Smoment[k]{-\frac{2}{3}} generally grows as a power law with \nparticle, resulting in a progressive shift towards weighting the outer regions of a halo as \nparticle\ increases, resulting in substantial mass-dependant biases).  For the work presented here, we use \nrank\eqx{-1} as our centrally weighted statistic throughout, which we have found to be a good compromise between the problems raised by \nrank\eqx{-1.5} and \nrank\eqx{-\frac{2}{3}}.  

For a given halo $\alpha$ with \nparticle\ particles, we will denote the maximum attainable value of \Smoment[k]{m} (\ie\ the case where all of a halo's particles are found in just one matching halo) as \Smomentmax{m}.  One practical advantage of choosing \nrank\eqx{-1} is that this maximum matching score is given by
\begin{equation}\label{eqn-max_score}
S^{{(}-1{)}}_{\rm{max}}(n_{\rm p}){=}\sum\limits_{r{=}1}^{n_{\rm p}} \frac{1}{r}
\end{equation}

\noindent which is the well known generating function of the Harmonic series, having the continuous approximation $S^{{(}{-1}{)}}_{\rm max}(n_{\rm p}){=}\gamma+\ln(n_{\rm p})$ where $\gamma{=}0.5772156649$ is the EulerÐ-Mascheroni constant.  In the trivial case of \nrank\eqx{0}, \Smomentmax{0} is simply equal to \nparticle.

For the optimally selected match for each halo, we will drop the subscript and denote the pseudo-moments computed for its match as \Smoment{m}.  We will find it useful to express the values of the pseudo-moments calculated for these matches using a normalizing \quotes{goodness of match} statistic \fmoment{m} given by the implicit relation
\begin{equation}\label{eqn-f_good_max_score}
S^{{(}{m}{)}}{=}S^{{(}m{)}}_{\rm max}(f^{{(}m{)}}_{\rm{g}}n_{\rm p})
\end{equation}

\begin{figure}
\begin{center}
\includegraphics[width=85mm]{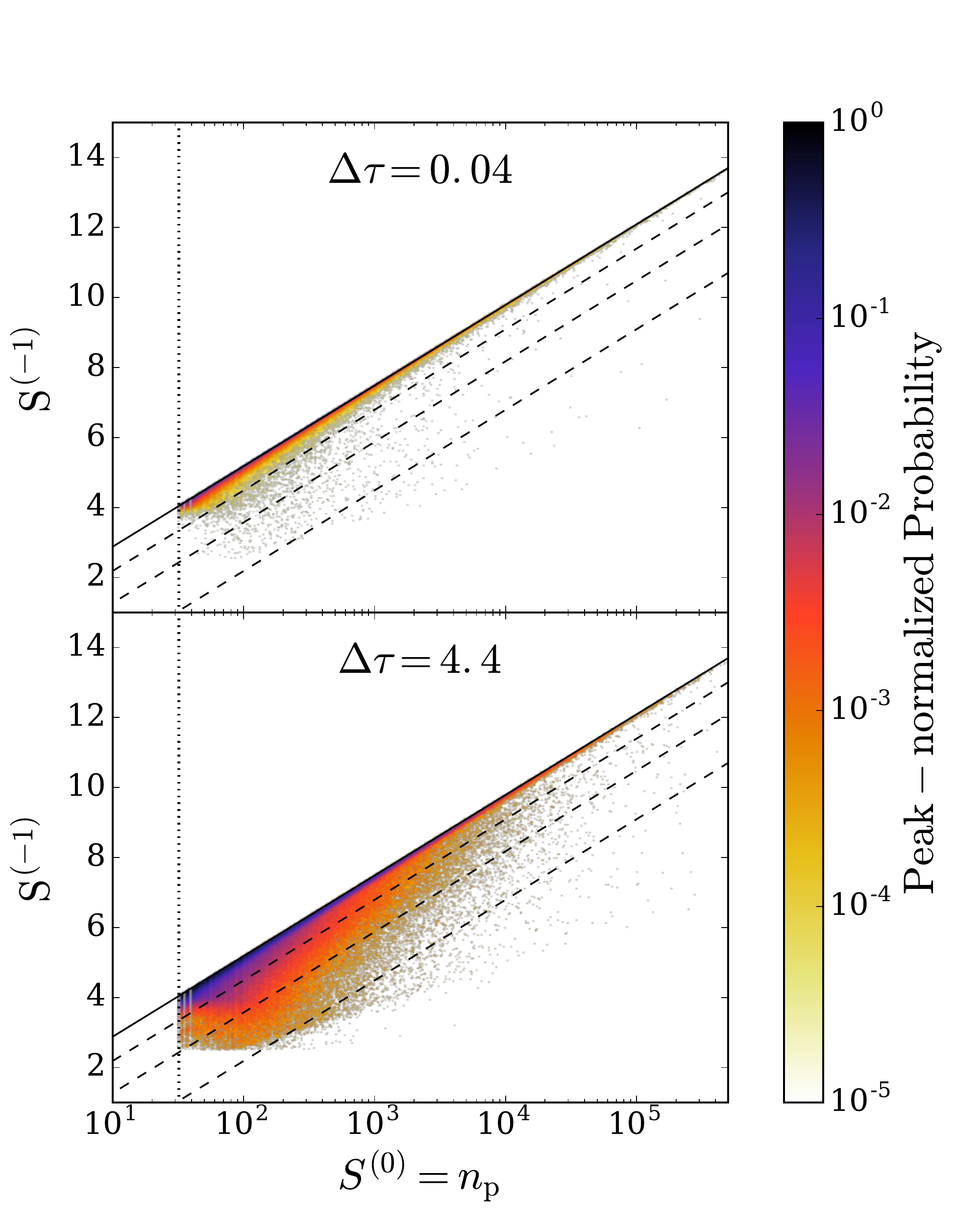}
\end{center}
\caption[Merger tree scores]{A figure illustrating the results of computing \Smoment{-1} for haloes of size \nparticle\ between two pairs of snapshots in the \GiggleZHR\ simulation separated by the shortest interval [top: \zx{j}\eqx{0} and \zx{i}\eqx{0.004} separated by \Deltatau\eqx{0.04} dynamical times] and by the longest interval [bottom: \zx{j}\eqx{0} and \zx{i}\eqx{0.56} separated by \Deltatau\eqx{4.4} dynamical times] considered in this paper.  The black solid line illustrates \Smomentmax{-1} (the maximum score attainable as a function of halo size) while the black dotted lines illustrate lines of constant \fmoment{-1}\eqx{50}, $20$ and $5$\% (ordered top to bottom respectively).  Vertical dotted lines are drawn at our minimum halo size of \nparticle${=}32$.\label{fig-match_scores}} 
\end{figure}

In Figure \ref{fig-match_scores} we illustrate the results of computing \Smoment{-1} between the haloes in two pairs of snapshots of the \GiggleZHR\ simulation: one pair at redshifts \zx{j}\eqx{0} and \zx{i}\eqx{0.004} and one pair at \zx{j}\eqx{0} and \zx{i}\eqx{0.56}.  These two snapshot pairs are separated by approximately \Deltatau\eqx{0.04}\ and \Deltatau\eqx{4.4} dynamical times respectively, covering the full range of temporal separations we consider in this present study.  Lines of constant \fmoment{-1}\eqx{50\%}, $20\%$ and $5\%$ are shown for comparison.

\subsubsection{Selecting \quotes{good} matches}

\begin{figure*}
\begin{minipage}{175mm}
\begin{center}
\includegraphics[width=125mm]{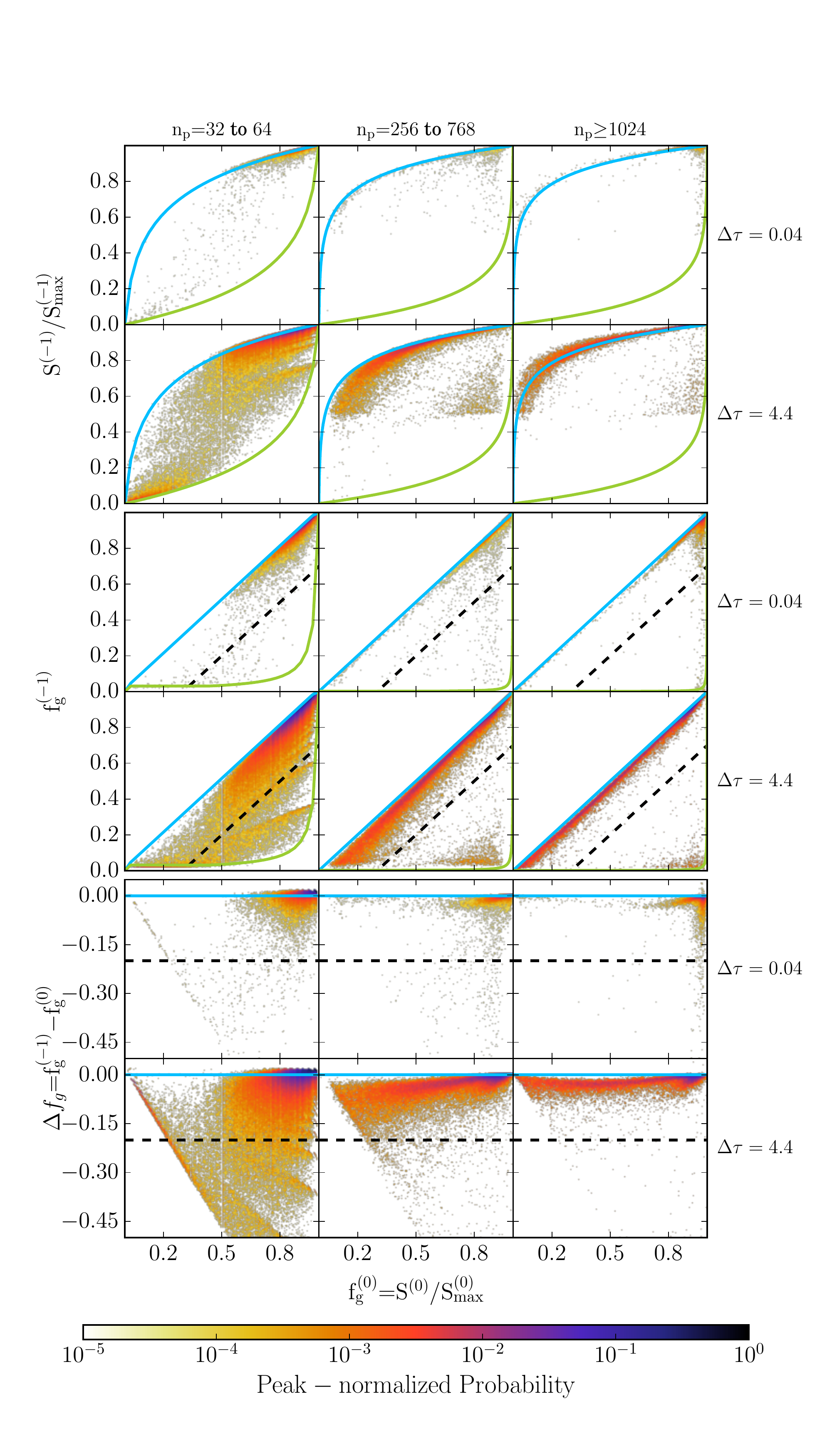}
\caption[Moment-moment plots]{A figure illustrating matching results in three representations of the \Smoment{-1}{--}\Smoment{0} space for three ranges of halo size ($32$ to $64$, $256$ to $768$ and ${\geq}1024$ particles) between two pairs of snapshots in the \GiggleZHR\ simulation separated by the shortest interval (\zx{j}\eqx{0} and \zx{i}\eqx{0.004} separated by \Deltatau\eqx{0.04} dynamical times) and by the longest interval (\zx{j}\eqx{0} and \zx{i}\eqx{0.56} separated by \Deltatau\eqx{4.4} dynamical times) considered in this paper.  Blue lines represent a maximal score model (left-to-right, for haloes with $64$, $768$ and $1024$ particles; note that the shape of these curves vary slightly with halo size, so some points for haloes with ${>}1024$ particles can be found above the maximal model plotted) whereby, for a given value of \fmoment{0}, all matching particles contiguously occupy the halo's \emph{lowest} ranks.  Green lines represent a minimal score model (left-to-right, for haloes with $32$, $256$ and $1024$) whereby, for a given value of \fmoment{0}, all matching particles contiguously occupy the halo's \emph{highest} ranks.  The black dashed line marks the $\Delta f_{\rm g}{=}-0.2$ cut we use to separate good matches (those on-or-above this line) from bad matches (those below this line).\label{fig-match_moments}}
\end{center}
\end{minipage}
\end{figure*}

There is an ambiguity in the \Smoment{-1} statistic that makes it susceptible to the problems introduced in Section \ref{sec-complicating_situations}: for a given halo match, is the value of \Smoment{-1} dominated by a small number of central particles, or a vast number of outer-halo particles?

Since \Smoment{0} is a measure of the number of particles involved in a match, we find that this ambiguity can be resolved when matches are examined in the \Smoment{-1}{--}\Smoment{0} plane.  In Figure \ref{fig-match_moments} we present matching results in this space for the two snapshot pairs illustrated in Figure \ref{fig-match_scores} (separated by \Deltatau\eqx{0.04} and $4.4$ dynamical times) for three halo size ranges ($32$ to $64$, $256$ to $768$ and ${\geq}1024$ particles).  While the situation is somewhat complicated for the smallest particle range (as a result of strong quantisation of \Smoment{-1} for such low particle counts), a clear bimodality exists in all cases.  Also plotted in this figure are two model lines representing minimal and maximal \Smoment{-1} models.  The maximal model (in blue) is the maximal \Smoment{-1} attainable for a given \Smoment{0} in the case where all particles involved in a match contiguously occupy the \emph{lowest} particle ranks of the halo's ID list.  The minimal model (in green) is the minimal \Smoment{-1} attainable for a given \Smoment{0} in the case where all particles involved in a match contiguously occupy the \emph{highest} particle ranks of the halo's ID list.  The halo matches can be seen to cluster along these two lines, with only those cases lying near the upper (blue) line representing cases where the \Smoment{-1} statistic could be dominated by material from the halo's core.

This bimodality can be accentuated and cases dominated by core material arrayed cleanly along a 1:1 line by plotting these matches (and models) in the \fmoment{-1}{--}\fmoment{0} space instead.  When we do this, poor matches dominated by the outer material of a halo are placed cleanly in the bottom right corner of this space, with core-dominated matches exhibiting a small scatter, on or slightly below the 1:1 line.  In the bottom-most panel of Figure \ref{fig-match_moments} we plot \Deltafgood${=}$\fmoment{-1}{--}\fmoment{0} along with a horizontal line illustrating the cut we use (see Section \ref{sec-convergence_quality_of_match} where we arrive at the value for this cut) in this space to separate matches that we will accept as \quotes{good} (those on-or-above the line) from matches that we reject as bad (those below the line).  When building matches we apply a preselection whereby we select only the good match (if possible) which passes our cut and maximises \Smoment{-1}.  Rejected matches are ignored for all tree-building purposes.

\subsubsection{Selecting \quotes{best} matches}\label{sec-good_matches}

The merger tree algorithm we present below will rely upon a process of scanning for matches -- both forwards and backwards in time -- to correct for occasions when structures are transiently absent from halo catalogs or become spuriously joined by a halo finder, only to separate again at later times.  We shall refer to instances where haloes \haloid{S}{ }{ } and \haloid{S}{ }{'} are matched from snapshot \ax{k} to \ax{j}${>}$\ax{k} as \quotes{forward} matches and cases where haloes \haloid{S}{ }{"} and \haloid{S}{ }{' } are matched from snapshot \ax{i} to \ax{j}${<}$\ax{i} as \quotes{back} matches.  In the simplest and most common cases, matching between haloes \haloid{S}{ }{ } and \haloid{S}{ }{'} is generally symmetric (\ie\ \haloid{S}{ }{ }\ is matched to \haloid{S}{ }{'} and \haloid{S}{ }{'} is matched to \haloid{S}{ }).  These cases will be said to be connected by \quotes{\twoway} matches.  In cases where this is not true, the match is said to be \quotes{\oneway}.

Additionally, because we will be searching for matches to-or-from a halo from-or-to a range of snapshots, matches may be found over a range of temporal intervals.  The case with the smallest interval will be referred to as the \quotes{most immediate} match.

Finally, throughout our algorithm we will seek to decide which match to consider the \quotes{best} to-or-from a halo.  To do so we will impose the following criteria:\\\\
\begin{tabular}{@{}ccl}
~& i.  &The best match will be the most immediate. \\
~& ii. & The best match will maximise the \Smoment{-1} statistic.
\end{tabular}\\

\begin{figure*}
\begin{minipage}{170mm}
\begin{center}
\includegraphics[width=165mm]{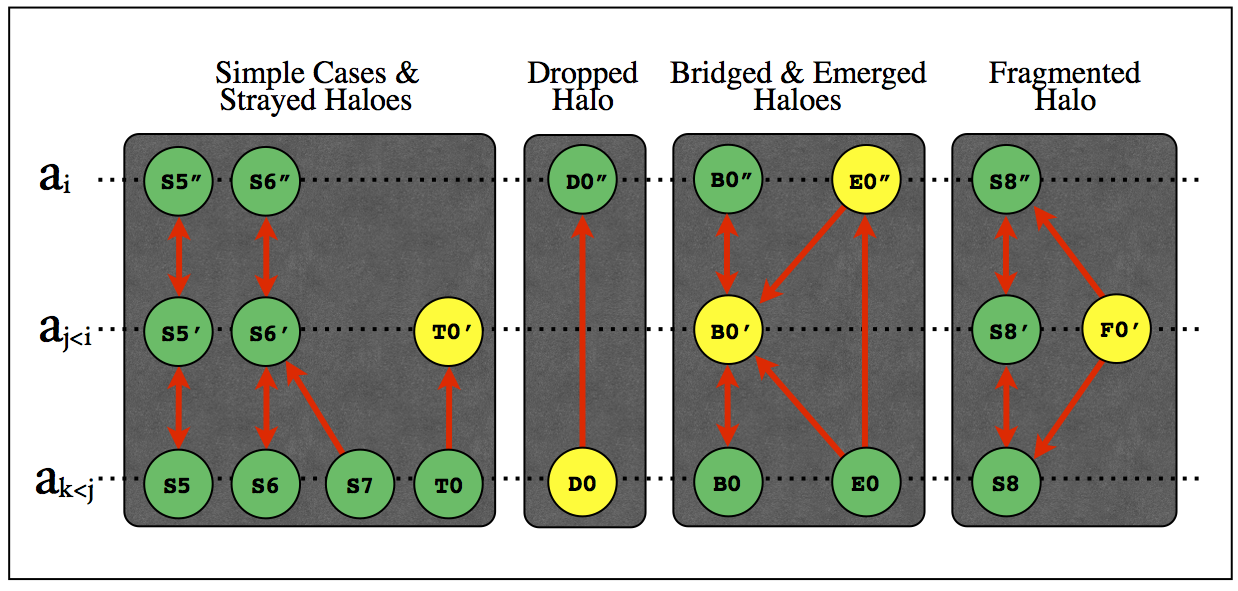}
\caption[Merger tree cases]{A figure illustrating the various cases - both normal and pathological - addressed by our merger tree algorithm.  Dotted horizontal lines represent progressive simulation snapshots \ax{i} where ${\rm i}$ runs from $0$ to \nsnap\mone\ in order of increasing expansion factor; ${\rm i{-}j}$ and ${\rm j{-}k}$ can be as large as the tuneable parameter \nsearch\ and vary for each individual case.  In \quotes{normal} cases, simple matches are made between every snapshot for every halo until it merges with its main progenitor line (as halo \haloid{S}{7}\ does with \haloid{S}{6}\ between snapshots \ax{k} and \ax{j}).  Occasionally, a halo is not successfully assigned a descendant (\eg\ \haloid{T}{0}{'}) and we refer to these and their progenitors as \quotes{strayed} haloes.  In the case of \quotes{dropped} haloes (\eg\ \haloid{D}{0}), the identity of the halo gets lost after snapshot \ax{k} but is recovered at a later snapshot \ax{i{>}k} (as \haloid{D}{0}{"}).  In the critical case of a \quotes{bridged} halo (\eg\ \haloid{B}{0}{'}), multiple back-matches get made to-or-from a halo with haloes at a later time (\ax{i{>}j}).  The back-matched halo identified as the best match to the bridged halo is identified as the bridge's main progenitor (\haloid{B}{0}{"}) while all other back-matched haloes become identified as candidate emerged haloes.  An emerged candidate that is successfully matched to by a halo at \ax{k{<}j} (\eg\ \haloid{E}{0}{"}) is identified as an \quotes{emerged} halo.  Emerged candidates which are not successfully matched to from a halo at \ax{k{<}j} (\eg\ \haloid{F}{0}{'}) are identified as \quotes{fragmented}.  
\label{fig-merger_trees_cases}} 
\end{center}
\end{minipage}
\end{figure*}

\subsection{Pathological cases defined}\label{sec-pathologies}

The process of halo finding can lead to the spurious disappearance, joining or splitting of haloes with important (sometimes cascading) negative consequences for merger trees.  In this subsection we detail our approach to classifying these artefacts and in the subsequent subsection we will present specific details of our algorithm for correcting for them when they occur.  Importantly, we choose an approach which is phenomenological in nature, describing these pathologies in terms of the network of matches involved rather than the physical processes responsible.  This approach should find equal applicability to any halo finder since artefacts of the type defined here are certain to be present at some level for all.  Specific results are likely to differ significantly however, with at least some artefacts likely to be significantly reduced for phase-space halo finders, for example.  

In Figure \ref{fig-merger_trees_cases} we present diagrams of several dysfunctional matching situations which lead to the classes of merger tree pathologies our algorithm is designed to address.  We shall refer to it throughout this subsection.

\subsubsection{Strayed haloes}

The left-most panel of Figure \ref{fig-merger_trees_cases} illustrates the typical cases of simple mergers, whereby matches between haloes are easy to determine and unambiguous (\halochainijk{S}{5}).  This can include mergers as well (\halochain{S}{7}{}{S}{6}{'}).

In cases where a good match can not be found between a halo and a descendant at a later snapshot, no future identity can be given to the halo and it (along with all progenitors) is said to be \emph{strayed} (\haloid{T}{0}{'} and \haloid{T}{0}{}).  This happens, for example, when small haloes are tidally disrupted and their particles are not subsequently incorporated into another halo or when small haloes are a result of numerical noise.  It can also result from inadequacies in the matching or tree building processes and as such, the abundance of strayed haloes is a useful diagnostic of our methodology.

\subsubsection{Dropped haloes}

A dropped halo occurs when a resolved halo -- which is accurately detected by the halo finder and tracked by the merger tree -- fails to be matched to another halo for one-or-more consecutive outputs (\halochainik{D}{0}).  This may occur, for example, for small isolated haloes at the limit of the halo finder's ability to accurately separate real structure from noise.

\subsubsection{Bridged and emerged haloes}

Bridged haloes occur when one-or-more haloes are momentarily joined by the halo finder (\haloid{B}{0}{'}), only to separate as distinct structures afterwards.  This may happen several times as haloes repeatedly pass through each other during the periodic apocentric passages of a merger event.  For configuration-space halo finders such as \Subfind, this occurs even for very large and well resolved haloes.  This situation is particularly disruptive to naive approaches of merger tree construction since failure to identify these occurrences will generally lead to labelling the system as a merger followed by the spontaneous creation of a (potentially very large) progenitor-less halo.

We refer to any haloes that separate from the main descendant line of a bridged halo as emerged haloes at the point where the halo finder is able to identify them as distinct substructures after being bridged (\haloid{E}{0}{"}; we do not refer to \haloid{B}{0}{"} as emerged however, since this halo is in the main progenitor line of \haloid{B}{0}).  This situation is similar to that of a dropped halo but with important differences: the progenitor of the emerged halo is clearly matched to a bridged halo during the intervening time (\halochain{E}{0}{}{B}{0}{'}), the emerged halo is clearly identified as subsequently emerging from that bridged halo's main progenitor line (\halochain{E}{0}{"}{B}{0}{'}), and a clear match is formed between the emerged halo and its progenitor (\halochainik{E}{0}).

Care must be taken in these cases to not only identify this situation, but to ensure the continuity of all involved halo identities across the snapshot(s) of the bridged halo event.  Incorrectly identifying the main progenitor line threading through the bridged halo event lasting from $a_{\rm i}$ to $a_{\rm k}$ can result in a failure to form the correct matches to emerged haloes or in spurious self-merger events, for example.  This requires a robust and consistent approach to matching and main progenitor selection.  These problems can be particularly aggravated by central-satellite switching events.  If main progenitor lines are not successful at robustly following the cores of haloes through these occurrences, situations can arise where a halo appears to emerge from itself.  Subsequent attempts to build progenitors to these events will effectively lead to spurious self-merging events.  This is one reason why our matching algorithm is optimised to follow the cores of haloes.

\subsubsection{Fragmented haloes}

With the main progenitor of a bridged halo determined and emerged haloes identified, the remaining haloes back-matched to a bridged halo are essentially analogous to what \citet[][FM08 henceforth]{Fakhouri:2008} refer to as \emph{fragmented} haloes.  However, we separate haloes described as fragmented by FM08 into two subclasses.  The first, are those which we have defined above as emerged haloes while the second are those which fail to have a progenitor assigned to them (\ie\ the connection \halochainik{E}{0} can not be made).  This may include real instances of halo fragmentation due to processes such as dynamical stripping (if the halo finder is able to detect such a structure), or be a result of a failure to build this match, either due to marginal matching statistics or failure to search over a long enough interval, for example.  In other words, fragmented haloes are those not assigned progenitors which would otherwise be considered emerged.

\subsection{The {\mdseries\gbpTrees} algorithm}\label{sec-gbpTrees}

To address the pathologies defined in the previous subsection, we have constructed an algorithm built upon a system of both forward and back-matches.  The use of back-matches ensures that we can identify cases where haloes appear to emerge from other haloes, allowing us to identify bridged, emerged and fragmented halo cases.  Trees for both FoF groups and substructures are constructed together simultaneously\footnote{Presently, substructure trees are constructed only for exclusive particle memberships, as presented in raw form by \Subfind.  Each substructure is tracked individually, with particles involved in sub-substructure not included in the matching process.} with roots initialised as the haloes of the last simulation snapshot ($i{=}{n_{\rm snap}{-}1}$).  For each snapshot previous to the last (${i}{<}{n_{\rm snap}{-}1}$), the following four steps are iteratively applied:
\\

\begin{figure*}
\begin{minipage}{170mm}
\begin{center}
\includegraphics[width=170mm]{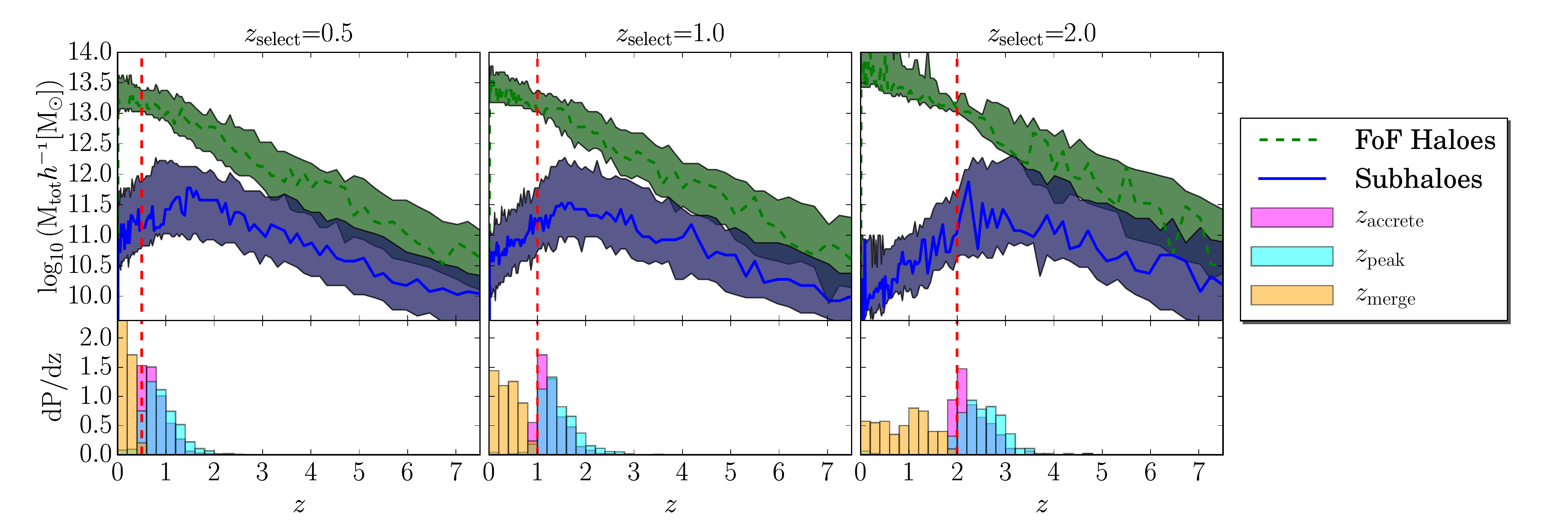}
\caption{The evolution of FoF masses [green] and their five most massive substructures [blue] (lines of respective colour identify peak probabilities of the distribution; shaded regions of respective colour identify the $68$\% confidence intervals) in \GiggleZHR\ for FoF systems more massive than $10^{13}$ \Msolhunit\ selected at redshifts $z{=}0.5$ [left], $z{=}1$ [middle] and $z{=}2$ [right] (identified with red vertical dashed lines).  Bottom panels illustrate the distributions of times when these substructures [cyan] reach their peak halo size, [magenta] are accreted onto their final FoF group and [orange] finally merge.\label{fig-M_tracks}}
\end{center}
\end{minipage}
\end{figure*}

\noindent \emph{i. Identify descendant-side matches}: For every halo, a list of matches to potential descendants is constructed.  We seek to ensure that we capture all halo-splitting events with this list.  Due to asymmetric matching situations (\ie\ cases where a halo matched to in the forward direction is not matched back to the same halo), this list must include all back matches to the halo as well as its best forward match.  For all contributions to this list, snapshots covering the range ${(}{i+1}{)}{\rightarrow}{(}{i{+}n_{\rm search}}{)}$ (but of course, only as far as the last snapshot of the simulation) are scanned.  Care is taken to keep only the most immediate matches of this list, with others removed by scanning along the descendant lines of each match.  Ultimately, all connections between nodes in the final merger trees are constructed from these lists.
\\

\noindent \emph{ii. Identify descendants}: For each halo, the best match (as defined in Section \ref{sec-good_matches}) from its list of descendant-side matches is selected as an initial preferred descendant.  Good matches to haloes in the snapshot range ${(}{i+1}{)}{\rightarrow}{(}{i{+}n_{\rm search}}{)}$ are then examined.  If one of these is found to be a candidate emerged halo (established in the following step; not possible in the first iteration) in the main-progenitor line of the currently preferred descendant, then this halo becomes the new preferred descendant.  This procedure is repeated for the full range of snapshots scanned, with matches to candidate emerged haloes in the (potentially changing) descendant line of the halo sought.  In this way, the algorithm is able to separate complex nested hierarchies of emerged haloes, identifying emerged haloes of emerged haloes, \etc
\\

\noindent \emph{iii. Identify bridges}: With descendants selected for all haloes in snapshot $i$, any haloes at snapshots $j{>}i$ involved in descendant-side matches to haloes in snapshot $i$ that have not been assigned a progenitor are marked as candidate emerged haloes.  Any haloes in the current snapshot with one-or-more haloes in its descendant-side match list labelled as a candidate emerged halo are marked as bridged haloes.
\\

\noindent \emph{iv. Finalise}: Because processing of subsequent snapshots may alter a snapshot's haloes' information (due to searches for emerged halo matches), snapshot $i$'s information is not written until snapshot $i{-}$\nsearch\ has been completely processed.  At that time, a finalisation step takes place.  Any candidate emerged haloes which have not been assigned a progenitor are marked as a new fragmented halo.  Any haloes not assigned a descendant are marked as strayed.  Furthermore, any haloes to which more than one progenitor-side match has been assigned but for which a progenitor has not been identified (this can happen if all progenitor-side matches are matched to emerged candidates afterwards, for example) has the best-matching progenitor-side match forced to be its progenitor.  With all descendant pointer decisions made, the current snapshot's haloes are added to the trees and results written to disk.
\\

Once the full range of snapshots have been processed in this way, another pass is performed, this time in reverse order from the earliest snapshot to the last to propagate important progenitor information up the trees.  This includes propagation of dominant halo identities (defined later), peak halo sizes and specification of primary and secondary haloes in merger events (all described in more detail later in Section \ref{sec-masses}), as well as communication of fragmented halo information along main progenitor lines (up to the point where a fragmented halo is found to merge with the halo it emerged from).

%% file: analysis.tex
In this section we apply the \gbpTrees\ algorithm to our suites of \GiggleZ\ and \Tiamat\ simulations.  We first explore how merger counts and the populations of pathological haloes defined in Section \ref{sec-pathologies} depend on our good match criterion (\Deltafgood), as well as the snapshot cadence (expressed in terms of the number of dynamical times separating snapshots, \Deltatausnap) and search interval (also expressed in units of dynamical time, \Deltatausearch) used to construct our merger trees.  With optimal choices for \Deltafgood, \Deltatausnap\ and \Deltatausearch\ established, we then illustrate the convergence of these populations with mass resolution and explore some properties of the pathological populations responsible for shaping these convergence behaviours.  We will finish our analysis by examining the merger rates of FoF haloes and subhaloes across cosmic time.

\subsection{Characterising Halo Mass}\label{sec-masses}

Before presenting the results of our convergence studies, a couple of important issues related to the characterisation of a substructure or FoF halo's mass must be addressed.  Summarising here: throughout this paper we use the instantaneous pre-merger particle count (corrected for sampling bias) of FoF haloes and the peak (inclusive, when possible) particle count of substructures to characterise halo masses.  This peak subhalo size calculation will require a robust approach to dealing with problems introduced by central-satellite switching, as we describe below.

We add that an alternative to using mass as our metric of halo size would be to use the peak of its circular velocity profile (\ie\ \Vmax).  Such a metric would be more robust to the problems we address here, but likely not immune.  As such, we expect our approach to be broadly useful for halo size metrics of all varieties.  Ultimately, we seek to compare several of our findings to previously published results expressed in terms of mass, so we use that characterisation throughout for this current work.

\subsubsection{Substructure}\label{sec-masses_substructure}

When characterising the mass of a subhalo, several options are available.  Choosing which to use depends ultimately on its intended application.  While a subhalo's instantaneous mass at any given time is a natural choice, there are several reasons why this is a significantly inappropriate choice for the majority of our current analysis.

To illustrate the point, we present in Figure \ref{fig-M_tracks} the mass evolution of the 5 most massive substructures of FoF groups in \GiggleZHR\ more massive than \TTTP{13}\Msolhunit\ at \zx{}\eqx{0.5}, $1$ and $2$.  The general patterns illustrated by this figure are generic to all redshifts and FoF halo masses resolved by our simulations.  Specifically, we see a monotonic increase of FoF mass throughout the simulation while substructures increase in mass only to a point (denoted the \quotes{peak halo size} point, occurring at redshift $z_p$) shortly before accretion onto their FoF group, with a monotonic decline in mass afterwards.  These trends \citep[which have been discussed previously in the literature; see][for example]{Behroozi:2014p2684} are not surprising of course: as subhaloes accrete onto larger systems, they will be subject to dynamical processes such as tidal stripping and will be unlikely to capture dynamically hot material from the massive system they merge into.  Also, \Subfind\ is designed to identify subhaloes out to the point where isodensity contours traverse a saddle point.  As a subhalo falls into a denser system, this isodensity contour naturally moves to higher densities, leading to a decline in the mass assigned to the substructure.  Both stripped material and lower density material lying outside this shifting contour will generally contribute to whatever system is considered the central of a relaxing system's potential, resulting in its growth at the expense of the merging substructure.

\begin{figure*}
\begin{center}
\begin{minipage}{175mm}
\includegraphics[width=160mm]{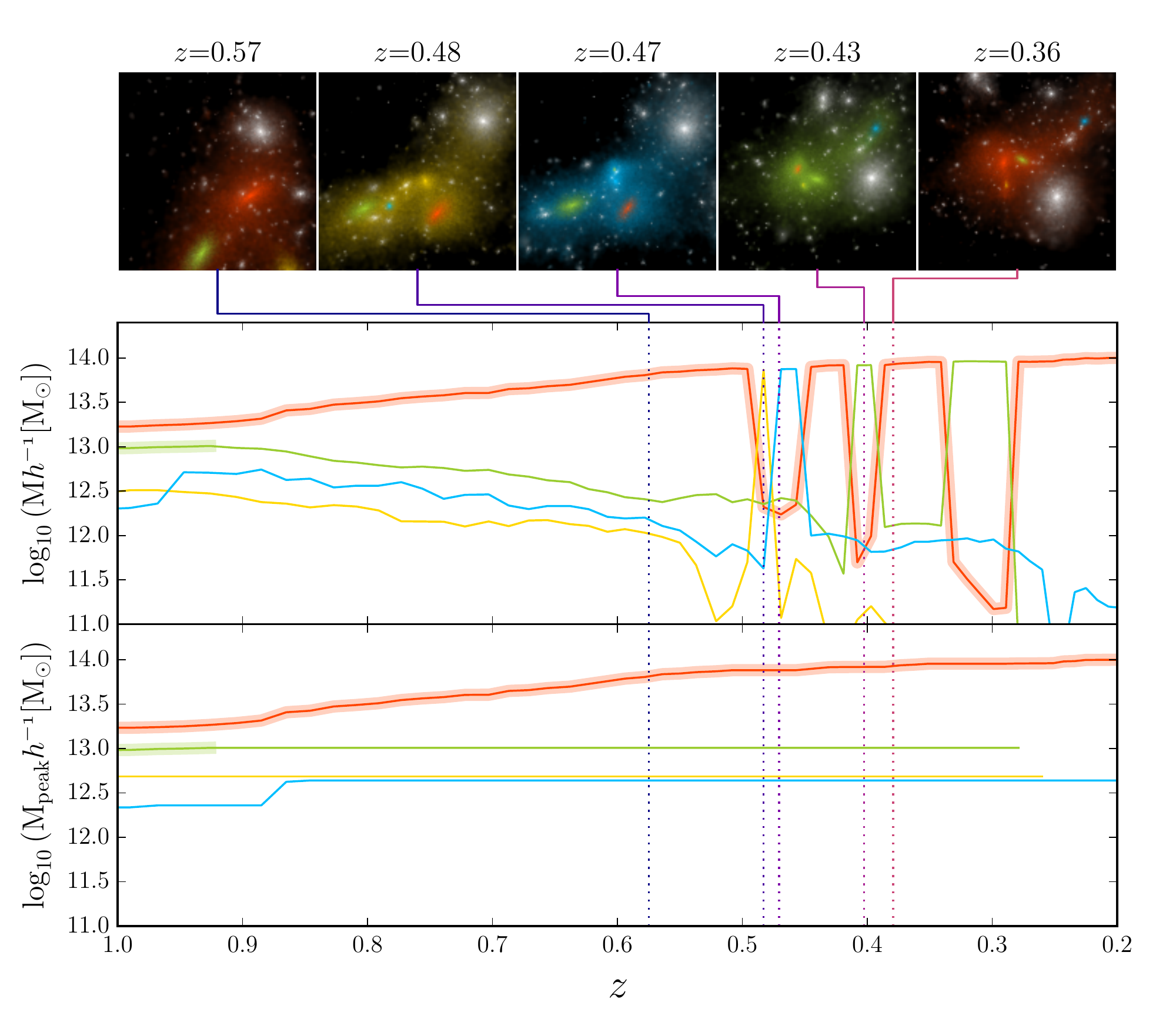}
\caption[Fiducial System]{An illustration of the evolution of the individual substructures involved in our \quotes{fiducial} merger case, extracted from the \GiggleZHR\ simulation, with which we illustrate several important concepts in this paper.  Individual substructures are colour coded consistantly throughout.  The evolution of substructure mass [top; with projected images -- $1.25$\Mpc\ on a side -- showing the structure of the system at a series of specific times shown above] and of peak substructure mass [bottom] are shown.  haloes identified as dominant are highlighted with thick lines.  Note that despite seven central-satellite switches ocuring during this event, the peak halo mass -- calculated using the dominant halo mechanism -- preserves the relative sizes of the individual substructures prior to merging.\label{fig-fiducial}} 
\end{minipage}
\end{center}
\end{figure*}

While an expected result, we offer this figure as an illustration of the hazards of simply using instantaneous mass for substructure merger tree comparisons, estimates of merger ratios, \etc\  The mass a subhalo has immediately prior to merging is going to depend significantly on a simulation's mass resolution and snapshot cadence, with finer resolutions in both being more likely to capture a subhalo at a point later in its history.  Furthermore, stellar mass will generally increase monotonically until the point when the subhalo reaches its peak mass.  Subsequent to this point, the relationship between subhalo mass and the galaxy it hosts diverges significantly.

For these reasons, we choose to characterise subhaloes by their peak mass throughout the analyses presented in this work.  This offers a fixed point that is identifiable to any simulation capable of resolving subhaloes to a given mass.  While short timescale mass fluctuations result in a slight upward bias of this measure with increasing snapshot densities, we have confirmed that this is negligible for our present analysis.  Several authors have found measures such as this to be an effective representation of galaxy halo properties in analysis such as subhalo abundance matching \citep{Conroy:2006p2686,Berrier:2006p2687}.

There is one highly significant problem for the application of this measure to substructures however: central-satellite switching.  The most straightforward approach to estimating the maximum size from a subhalo's progenitor line is often substantially disrupted by these events in our \Subfind\ catalogs.  We resolve this problem by introducing the concept of a \quotes{dominant} subhalo.  The dominant subhalo of a FoF group is meant to follow the descendant line which spends the longest amount of time as its central subhalo.  When computing peak masses for substructures, we only consider instances when the subhalo is a central halo in the calculation if it is the FoF group's dominant halo.  While this may lead to the occasional instance when a FoF group's central halo does not participate in the peak subhalo mass of any substructure, the result is a highly effective suppression of peak mass estimate biases due to central-satellite switching.

The dominant subhaloes of FoF groups are established during the progenitor propagation stage of our merger tree construction.  At the leaves of the merger trees of our FoF groups -- when FoF haloes are first resolved in our simulations -- the most massive subhalo of the system is initialised as its dominant subhalo.  We then propagate these identities forward along the descendant line of the subhalo until the host FoF group is found to merge with another FoF group.  At this point, only the dominant subhalo of the most massive of the merging FoF groups retains this designation, ensuring that only one subhalo in a FoF group is designated as dominant at any time.  Care must be taken when gaps in the dominant halo's progenitor line occur, resulting in rare instances when a FoF group is not assigned a dominant halo.  The fact that we build merger trees for both the FoF and substructure haloes of our catalogs in parallel facilitates this process.

In Figure \ref{fig-fiducial} we present the results of applying this approach to the fiducial system introduced earlier in Section \ref{sec-complicating_situations}.  The mass accretion histories of subhaloes which are designated as dominant are highlighted with thick lines.  Upon approaching this merging system, the yellow, blue and green subhaloes were obviously secondary components of the merger, being significantly smaller than the red subhalo.  Despite spending time as the central of the system (twice, in the case of the green subhalo) the peak size that we calculate for these subhaloes continues to preserve this fact, providing an effective characterisation of their pre-merger mass.  Using the peak masses of two merging systems as found at the time the secondary (\ie\ smaller) system reaches its peak size for the calculation of merger mass ratios, \etc\ provides a metric which more faithfully characterises the event.

An additional complication when comparing subhaloes across simulations with varying mass resolutions is introduced by the fact that increases in mass resolution drive increases in the fraction of a halo resolved as substructure.  As a result, the mass reported by \Subfind\ for one identical subhalo simulated at several mass resolutions will decline systematically with improved mass resolution.  In cases where we report final results or perform comparisons across various mass resolutions, we correct for this by using the substructure hierarchy pointers reported by \Subfind\ to report \quotes{inclusive} subhalo masses which include the mass of a substructure and of all substructures reported by \Subfind\ to be connected to it.  Substructure hierarchy pointers are not available for the \GiggleZ\ simulations, but are available for all others.

\subsubsection{Friends-of-friends Haloes}\label{sec-masses_substructure}

In the case of FoF haloes, we use a distinctly different strategy for estimating mass.  We find that transient mass spikes occur due to halo bridging for FoF haloes, but without recourse to the dominant halo mechanism we can not utilise peak mass estimates for these haloes.  Fortunately, FoF masses tend to monotonically increase until they merge, so the instantaneous particle count prior to merging is a reasonably robust characterisation of its peak size.  It should be noted that this is not strictly true -- haloes classified as emerged shortly before merging can be significantly smaller than their peak size at the latest stages of their evolution -- but such instances are generally uncommon.  This is the approach taken by others in the literature and facilitates comparison to previous studies.  

Several published studies have explored the issues of resolution-driven biases in FoF halo mass estimates due to sampling bias \citep[\eg][]{Warren:2006p2669,Lukic:2009dt} or systematic trends in the accuracy with which simulations treat nonlinear collapse \citep[\eg][]{2010ApJ...711.1198T}.  We choose to correct for sampling bias by adopting the FoF halo mass correction published by \citet{Warren:2006p2669}:
\begin{equation}\label{eqn-Warren_correction}
{n_{\rm p}'}{=}n_{\rm p}\left(1-n_{\rm p}^{-0.6}\right)
\end{equation}

\noindent and use this throughout when characterising FoF halo masses.  We note however that results in the literature differ on this subject -- even qualitatively for halo sizes \nparticle${<}{100}$ -- where \citet{Warren:2006p2669} report that sampling bias \emph{overestimates} FoF halo masses while \citet{2010ApJ...711.1198T} report that differences in non-linear collapse result in a net \emph{suppression} of FoF halo masses.  However, in the regime \nparticle${>}75$ for which we restrict analysis for all fits and quantitative analysis reported later, the magnitude of this disagreement is small.  However, anecdotally we note that the adoption of this correction ultimately improves the convergence of merger rates with mass resolution presented later in this work.  This suggests that merger rates may provide an interesting avenue from which to approach issues of halo mass biases in future studies.

\subsection{Convergence}\label{sec-convergence}

In this subsection we seek to demonstrate -- for both FoF and substructure haloes -- the convergence properties of the \gbpTrees\ algorithm using the mass dependence of total merger counts (specifically, the total number of times a halo above a given mass participates as a secondary system in a merger) and of pathology population sizes (specifically, the probability that a halo above a given mass lies in the progenitor line of a strayed or fragmented halo).  Total integrated merger rates are the metric of primary interest because they are a measure of the total number of branches in a set of trees, and hence of the total number of potential galaxies involved in a semi-analytic galaxy formation calculation.  Spurious creation of galaxies is one of the consequences of tree pathologies that we are trying to avoid, and so this metric will reflect the degree to which we are achieving this.  In terms of tree pathologies, we focus our attention on the probabilities that a halo above a given instantaneous size exists in the progenitor line of a halo that ultimately becomes strayed or is a descendant of a halo marked as fragmented.  Haloes marked as fragmented and found to be a progenitor of a halo with multiple progenitors are not considered mergers.

Note that for our studies of convergence with the quality of match criterion; search interval; and snapshot cadence, masses reported for substructures will not be inclusive of their own substructure (since substructure hierarchy pointers are not available for the \GiggleZ\ simulations).  For our study of mass resolution convergence, substructure masses are inclusive of their substructure (necessary, since significant mass-resolution dependant biases would be introduced otherwise).

\subsubsection{Quality of match criterion}\label{sec-convergence_quality_of_match}

\begin{figure*}
\begin{center}
\begin{minipage}{175mm}
\vspace{5mm}
\includegraphics[width=175mm]{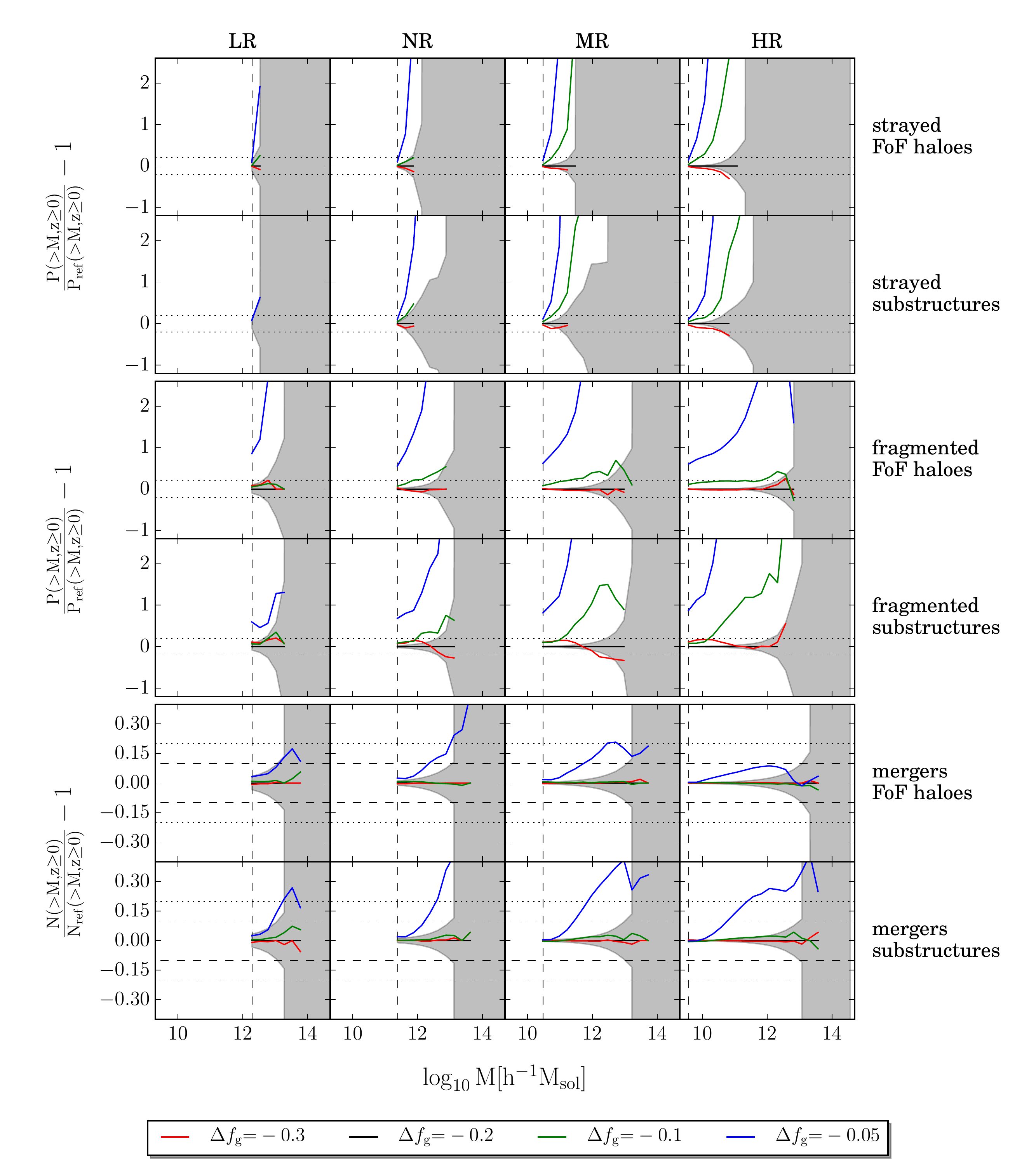}
\caption[Quality of match criterion convergence]{Three panels illustrating how the goodness of match criterion (\Deltafgood) affects the probability of a halo being identified as pathological (strayed [top panel] or fragmented [middle panel]) and the total number of mergers [bottom panel] for both FoF haloes [top in each panel] and substructures [bottom in each panel].  All quantities are averaged or integrated for redshifts $z{>}0$.  Grey shaded regions illustrate Poisson uncertainties for the fiducial case.  Each column illustrates one of the 4 \GiggleZ\ control volume simulations (increasing in mass resolution from \GiggleZLR\ [left], to \GiggleZNR\ [left-centre], to \GiggleZMR\ [right-centre] to \GiggleZHR\ [right]; our minimum halo size of \nparticle${=}32$ is illustrated with a vertical dashed line in each case) with colours illustrating fractional differences between results obtained using our preferred value (\Deltafgood${=}-0.2$; the black reference in all cases) and various values less strict (\Deltafgood${=}-0.3$ in red) or more strict (\Deltafgood\eqx{-0.05} in blue; \Deltafgood\eqx{-0.1} in green) than this.\label{fig-convergence_goodness}} 
\end{minipage}
\end{center}
\end{figure*}

\begin{figure*}
\begin{center}
\begin{minipage}{175mm}
\vspace{5mm}
\includegraphics[width=175mm]{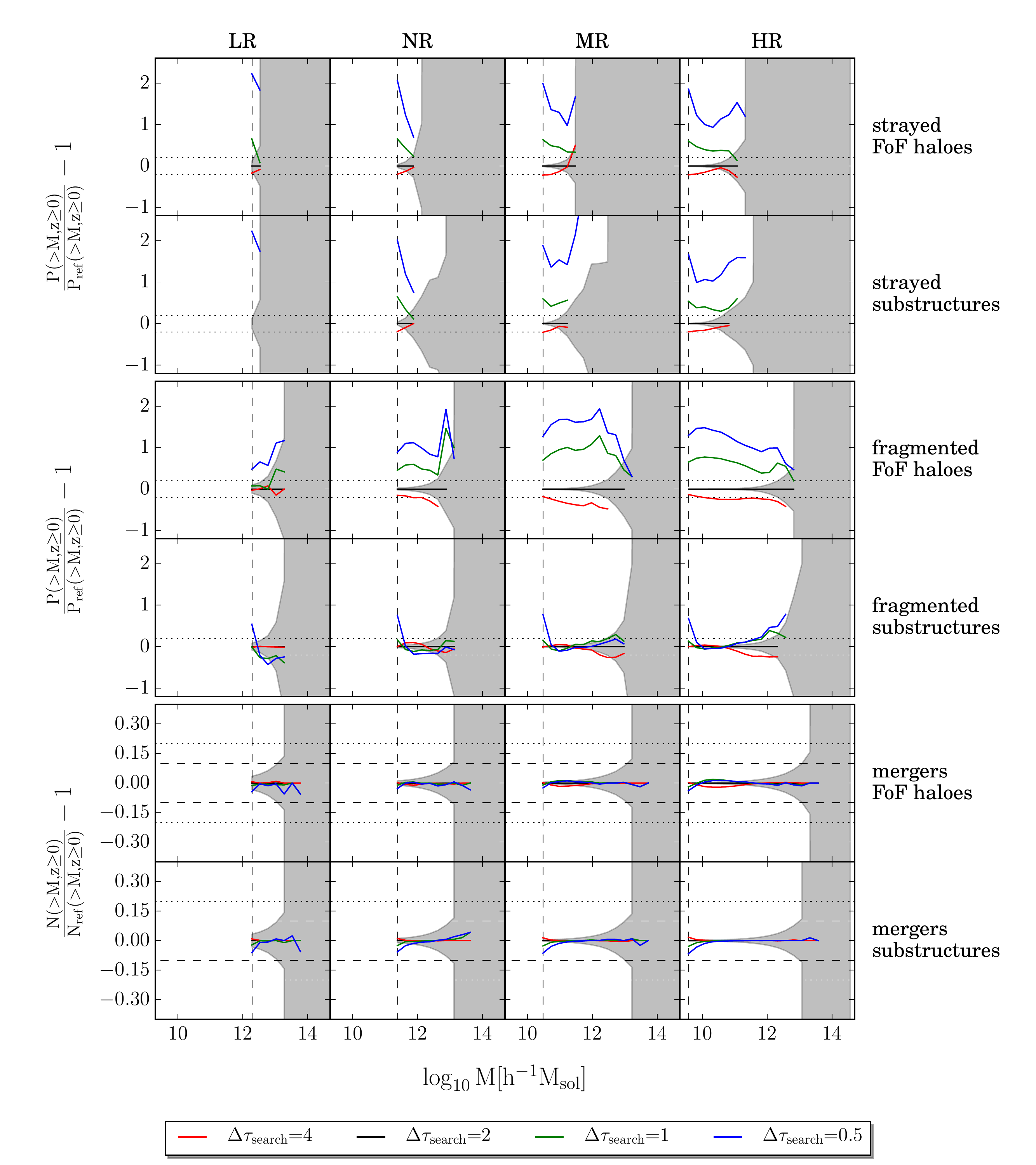}
\caption[Scan interval convergence]{Three panels illustrating how the interval over which matches are scanned to correct tree pathologies (\Deltatausearch; in units of dynamical times) affects the probability of a halo being identified as pathological (strayed [top panel] or fragmented [middle panel]) and the total number of mergers [bottom panel] for both FoF haloes [top in each panel] and substructures [bottom in each panel].  All quantities are averaged or integrated for redshifts $z{>}0$.  Grey shaded regions illustrate Poisson uncertainties for the fiducial case.  Each column illustrates one of the 4 \GiggleZ\ control volume simulations (increasing in mass resolution from \GiggleZLR\ [left], to \GiggleZNR\ [left-centre], to \GiggleZMR\ [right-centre] to \GiggleZHR\ [right]; our minimum halo size of \nparticle${=}32$ is illustrated with a vertical dashed line in each case) with colours illustrating fractional differences between results obtained using our preferred value (\Deltatausearch\eqx{2}; the black reference in all cases) and various values longer (\Deltatausearch\eqx{4} in red) or shorter (\Deltatausearch\eqx{0.5} in blue; \Deltatausearch\eqx{1} in green) than this.\label{fig-convergence_n_search}} 
\end{minipage}
\end{center}
\end{figure*}

\begin{figure*}
\begin{center}
\begin{minipage}{175mm}
\vspace{5mm}
\includegraphics[width=175mm]{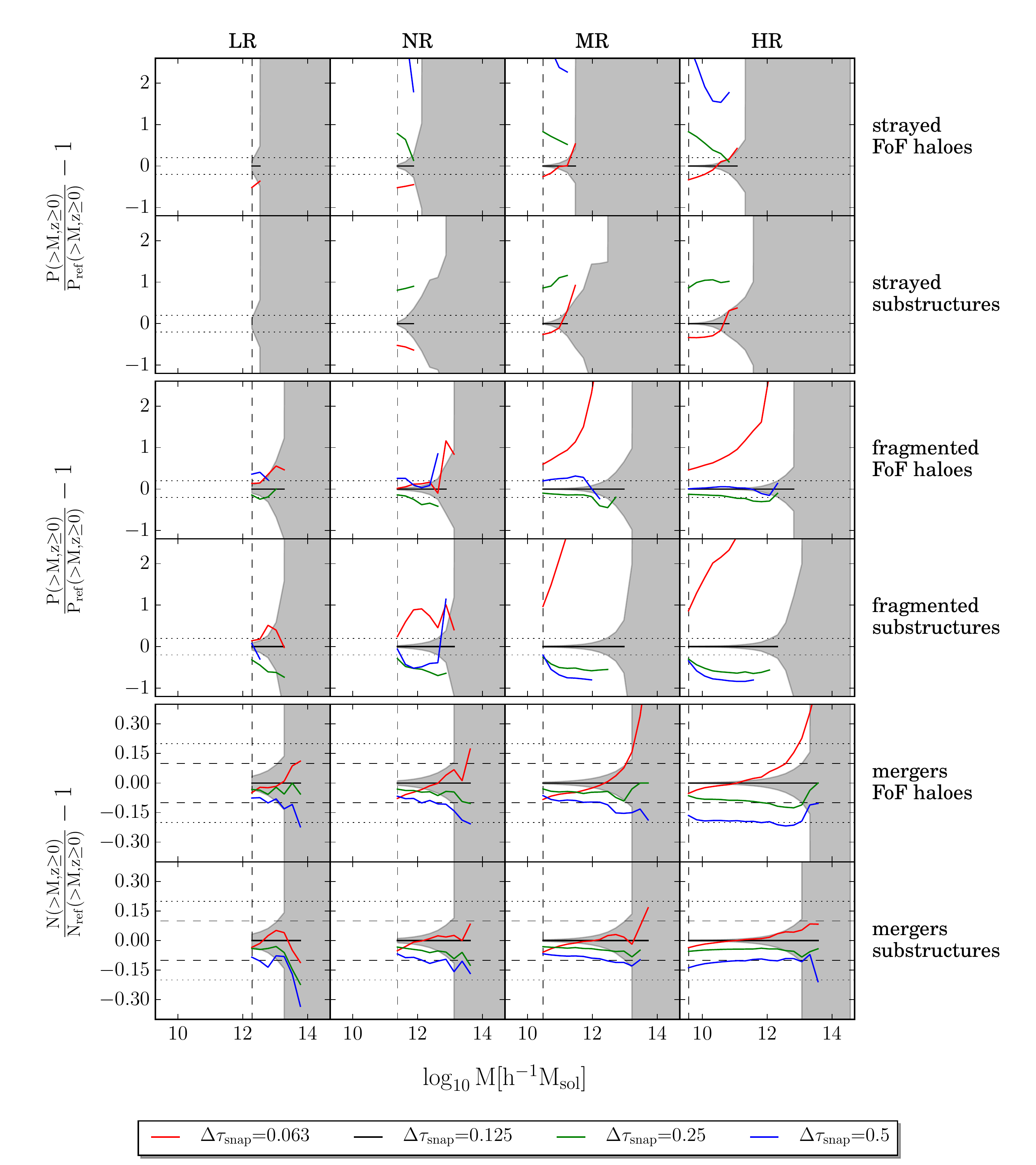}
\caption[Snapshot cadence convergence]{Three panels illustrating how the interval between snapshots (\Deltatausnap; in units of dynamical times) affects the probability of a halo being identified as pathological (strayed [top panel] or fragmented [middle panel]) and the total number of mergers [bottom panel] for both FoF haloes [top in each panel] and substructures [bottom in each panel].  All quantities are averaged or integrated for redshifts $z{>}0$.  Grey shaded regions illustrate Poisson uncertainties for the fiducial case.  Each column illustrates one of the 4 \GiggleZ\ control volume simulations (increasing in mass resolution from \GiggleZLR\ [left], to \GiggleZNR\ [left-centre], to \GiggleZMR\ [right-centre] to \GiggleZHR\ [right]; our minimum halo size of \nparticle${=}32$ is illustrated with a vertical dashed line in each case) with colours illustrating fractional differences between results obtained using our preferred value (\Deltatausnap\eqx{0.125}; the black reference in all cases) and various values more dense (\Deltatausnap\eqx{0.063} in red) or less dense (\Deltatausnap\eqx{0.5} in blue; \Deltatausnap\eqx{0.25} in green) than this.\label{fig-convergence_n_snap}} 
\end{minipage}
\end{center}
\end{figure*}

The quality of match criterion was introduced in Section \ref{sec-good_matches} as a parameter which establishes whether a halo match should be considered good or bad (and hence, ignored for all tree building purposes).  If insufficiently strict, tree pathologies will proliferate.  If too strict, matches which should be considered good will be rejected resulting in larger populations of strayed and fragmented haloes with consequent effects on merger rates.  As such, our goal is to establish the strictest \Deltafgood\ cut which does not show a substantial change in these populations.  

In Figure \ref{fig-convergence_goodness} we demonstrate the changes in the populations of strayed, fragmented and merging FoF and substructure haloes as we vary \Deltafgood\ between values of $-0.05$ and $-0.3$.  Observed trends are as expected: the highest values of \Deltafgood\ (\ie\ the most aggressive cuts) result in very large populations of tree pathologies and mergers.  Merger counts are stable at a level of \lsimx{1}\% for \Deltafgood\lsimx{-0.2} while populations of strayed and fragmented haloes are typically stable at a level of \lsimx{10}\% level.  These are both comparable to or less than the level of convergence found for the scan interval and snapshot cadence (see below) and as such, are sufficient for our present purposes.  We adopt a value of \Deltafgood\eqx{-0.2} throughout the rest of this paper.

\subsubsection{Scan interval}\label{sec-convergence_search}

\begin{figure*}
\begin{minipage}{170mm}
\begin{center}
\includegraphics[width=170mm]{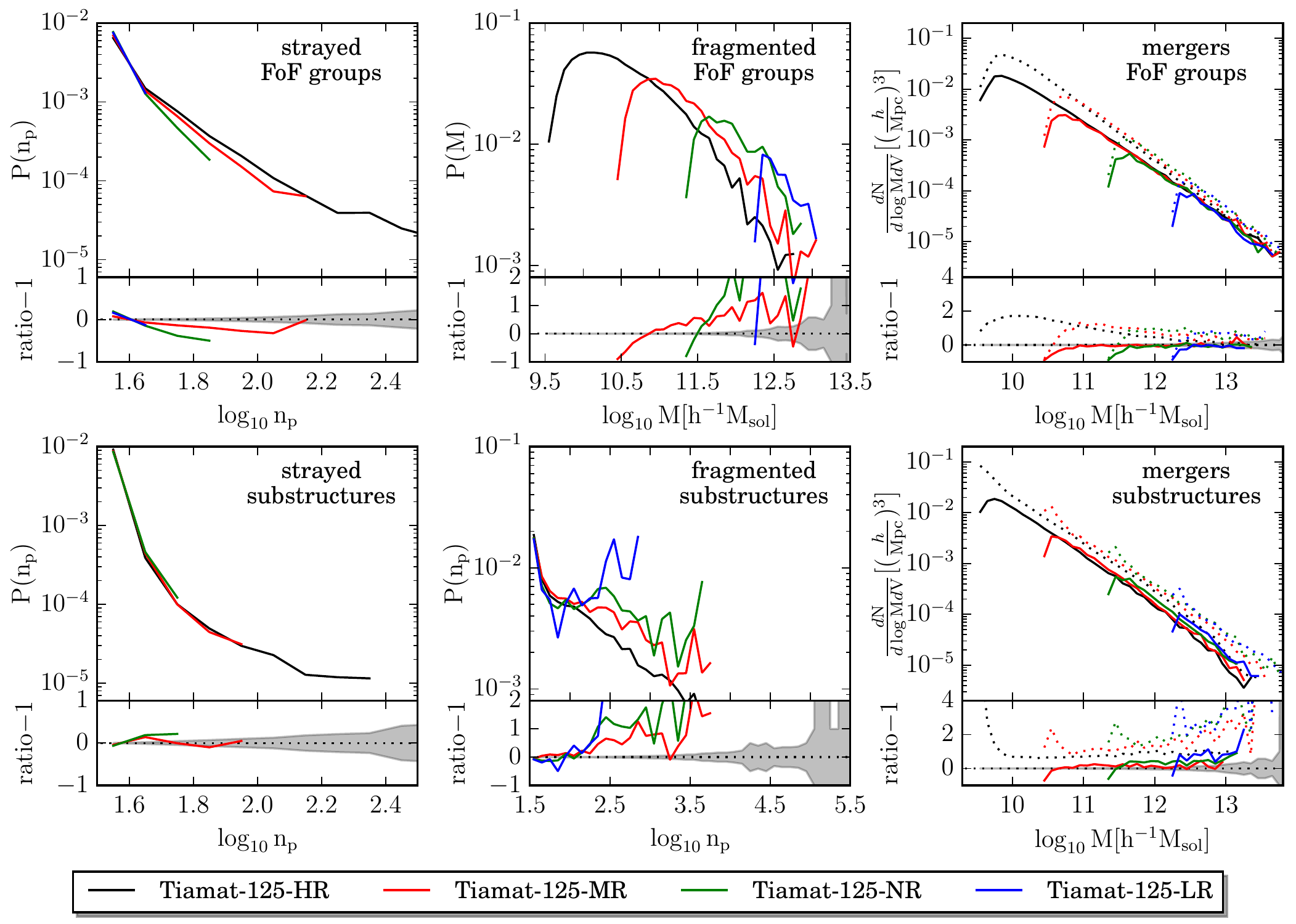}
\caption{The probabilities of haloes being identified as strayed [left] or fragmented [middle] and the density of mergers [right] as a function of mass or particle number (\nparticle) for both FoF groups [top] and substructures [bottom], averaged over the redshift interval $0{\leq}z{\leq}0.5$.  In all cases, results are shown for trees built with our nominal parameters [solid] and for the case where bridge fixing is turned off [dashed, mergers only].  Each of the 4 GiggleZ control volume simulations are illustrated with varying colours (increasing in mass resolution from \GiggleZLR\ [blue], to \GiggleZNR\ [green], to \GiggleZMR\ [red] to \GiggleZHR\ [black]).  Bottom panels in each case show the fractional difference of each from the \GiggleZHR\ results.  Note the variations in ordinate between plots, chosen to highlight the most relevant dependencies on \nparticle\ or mass.\label{fig-pops}}
\end{center}
\end{minipage}
\end{figure*}

The scan interval (\Deltatausearch) is the period over which matches are examined for the purposes of correcting tree pathologies.  By increasing \Deltatausearch, we increase the chances of identifying the descendant of a halo which has gone missing to a halo finder for a period of time.  Larger values of \Deltatausearch\ should lead to reduced populations of pathological haloes, but at the expense of increases to the runtime wallclock, storage and RAM resources required to generate the trees.  As a result, we seek the minimal value of \Deltatausearch\ which demonstrates convergence, in order to minimise resource requirements for future work.

In Figure \ref{fig-convergence_n_search} we find that expected trends are observed.  As \Deltatausearch\ is increased, we find a progressive decline in strayed haloes for both FoF haloes and substructures as the cores of long-lost haloes become detectable once-again to \Subfind.  The incidence of strayed haloes with \nparticle\gsimx{300} of both halo types drops to essentially zero as \Deltatausearch\ increases to ${\sim}2$ dynamical times with convergence at the ${\sim}20$\% level for haloes smaller than this.

The behaviour of the fragmented halo population follows a very similar trend for FoF groups, reaching convergence at a ${\sim}20$\% level over nearly the entire mass range.  The fragmented subhalo population is much less dependant on \Deltatausearch\ however, reaching a similar level of convergence (even better for \nparticle\lsimx{300}) by \Deltatausearch\eqx{1}.  

For merger rates, we find that even a short scan interval of \Deltatausearch\eqx{0.5} is adequate to secure merger rate convergence of \lsimx{1}\% for all but the smallest (\nparticle\lsimx{100}) haloes.  Even for these haloes, convergence at this level is secured for both halo populations by \Deltatausearch\eqx{2}.  The insensitivity of merger rates to \Deltatausearch\ reflects the effectiveness by which our algorithm prevents the creation of (or at least, classifies and isolates) spurious progenitor lines, even for very short search intervals (albeit, at the expense of designating a higher incidence of haloes as pathological in such a case).

\subsubsection{Snapshot cadence}\label{sec-convergence_snap}

Snapshot cadence refers to the temporal density of snapshots produced by the simulation and used to construct a set of merger trees.  One might naively expect that the use of more snapshots would naturally lead to more precise merger trees, given the additional information available for correcting pathologies \etc\  Unfortunately, the variations in merger counts and pathology populations with \Deltatausnap\ presented in Figure \ref{fig-convergence_n_snap} indicate that the situation is much more complicated.

For strayed haloes, convergence with \Deltatausnap\ is qualitatively similar to that with \Deltatausearch, with a level of \simx{20}\% reached by \Deltatausnap\eqx{0.125}.  Convergence of fragmented haloes (both FoF and substructure) is much more complicated however.  For FoF haloes, fragmented populations appear to be converging as \Deltatausnap\eqx{0.25} is reached.  At lower values, the population increases abruptly and substantially.  A similar trend is found for subhaloes, although this transition occurs at the larger value of \Deltatausnap\eqx{0.5}.  It is unclear at this point what exactly is driving this behaviour, but initial examinations indicate that it is a result of rare and very brief moments when haloes reach states where good matches can not be formed between an emerged halo and its correct progenitor.  By decreasing \Deltatausnap, the chances of a halo being caught in such a state at some point in its progenitor line increase, leading to an increase in fragmented halo populations.

Despite this, the convergence of merger counts is reasonably well behaved.  For \Deltatausnap\eqx{0.125}, they are converged at a level of approximately \lsimx{5}\% for all but the highest masses where convergence is at a level of $10$ to $20$\%, although this is only marginally Poisson-significant and occurs in a regime where systematics in mass-assignment have the greatest effect.  It is unclear to what degree errors in progenitor lines are responsible for this apparently reduced performance at high masses.

Conveniently, \Deltatausnap\eqx{0.125} is very similar to the snapshot cadence \citet{Benson:2012p2682} finds to be necessary for 5\%-level convergence of galaxy masses in semi-analytic galaxy formation modelling.

\subsubsection{Mass resolution}\label{sec-convergence_mass_res}

Having established a nominal set of parameters for our tree building algorithm, (\Deltafgood,\Deltatausearch, \Deltatausnap)${=}$($-0.2$,$2$,$0.128$), we look now at how the mass resolution of a simulation affects merger counts and pathological halo populations.  In Figure \ref{fig-pops} we present these populations for the \Tiamatcontrol\ simulations.  We find excellent mass resolution convergence for both merger counts and strayed halo populations.

At our highest mass resolutions, populations of fragmented haloes are converging, but much more slowly.  For FoF haloes, a level of convergence of ${\sim}50$\% over most mass ranges is found while convergence of fragmented substructures occurs instead in particle count, converging towards a declining power law with convergence at a level of \lsimx{50}\% for haloes with \lsimx{300} particles; much poorer for larger (albeit, much rarer) cases.  The fact that convergence for fragmented populations occurs by mass for FoF groups suggests a physical origin (\eg\ the ejection of substructures as new FoF groups from disrupted systems after a time exceeding the scanning interval, subsequent to their initial accretion) while the convergence of fragmented subhaloes by particle count suggests a numerical origin (\eg\ artefacts of inadequacies of the tree-building algorithm).  Regardless, these populations are small and confined to small systems.

Importantly, merger rates are converged at an excellent level for both cases.  We find that the total number of FoF mergers as a function of mass is converged for all mass resolutions at a level of ${\sim}5$\% when \nparticle${\gtrsim}75$.  For substructures, convergence with mass resolution is less complete but approaches a level of ${\sim}10$\% for \nparticle${>}100$ at mass resolutions $m_{\rm p}{\lsim}{10}^{9}$ \Msolhunit.  

Finally, also shown with dotted lines in this figure is the merger rates that result when our bridged and emerged halo finding apparatus is turned off.  We find that when care is not taken to identify and correct for tree pathologies, FoF merger rates are typically biased up by $50$ to $100$\% while substructure merger rates are generally biased up by ${>}100$\%.  These numbers are appropriate for our nominal tree-building parameters and may be much worse in other situations, particularly at denser snapshot cadences where there is a larger cumulative probability for tree pathologies to be introduced.

\subsection{Pathology analysis}\label{sec-pathology_analysis}

In Section \ref{sec-convergence_search} we found that fragmented halo and merger rates converge more quickly with \Deltatausearch\ for subhaloes than for FoF haloes. These results imply that the period of time for which substructures drop-out of the \Subfind\ catalogs is much shorter than the periods of time over which FoF haloes become lost to the friends-of-friends algorithm.  The essential difference is the higher densities reached by the isodensity contours effectively defining the boundaries of subhaloes versus those that effectively define the boundaries of FoF haloes.  As a result, the background densities within which subhaloes tend to get lost is much higher than those where FoF haloes lose their identities.  The dynamical times of the environments they get lost in are shorter as a result, leading to quicker evolution to their later state of detectability.  In this section, we look at this difference in more detail.

\subsubsection{Timescales}

\begin{figure}
\includegraphics[width=84mm]{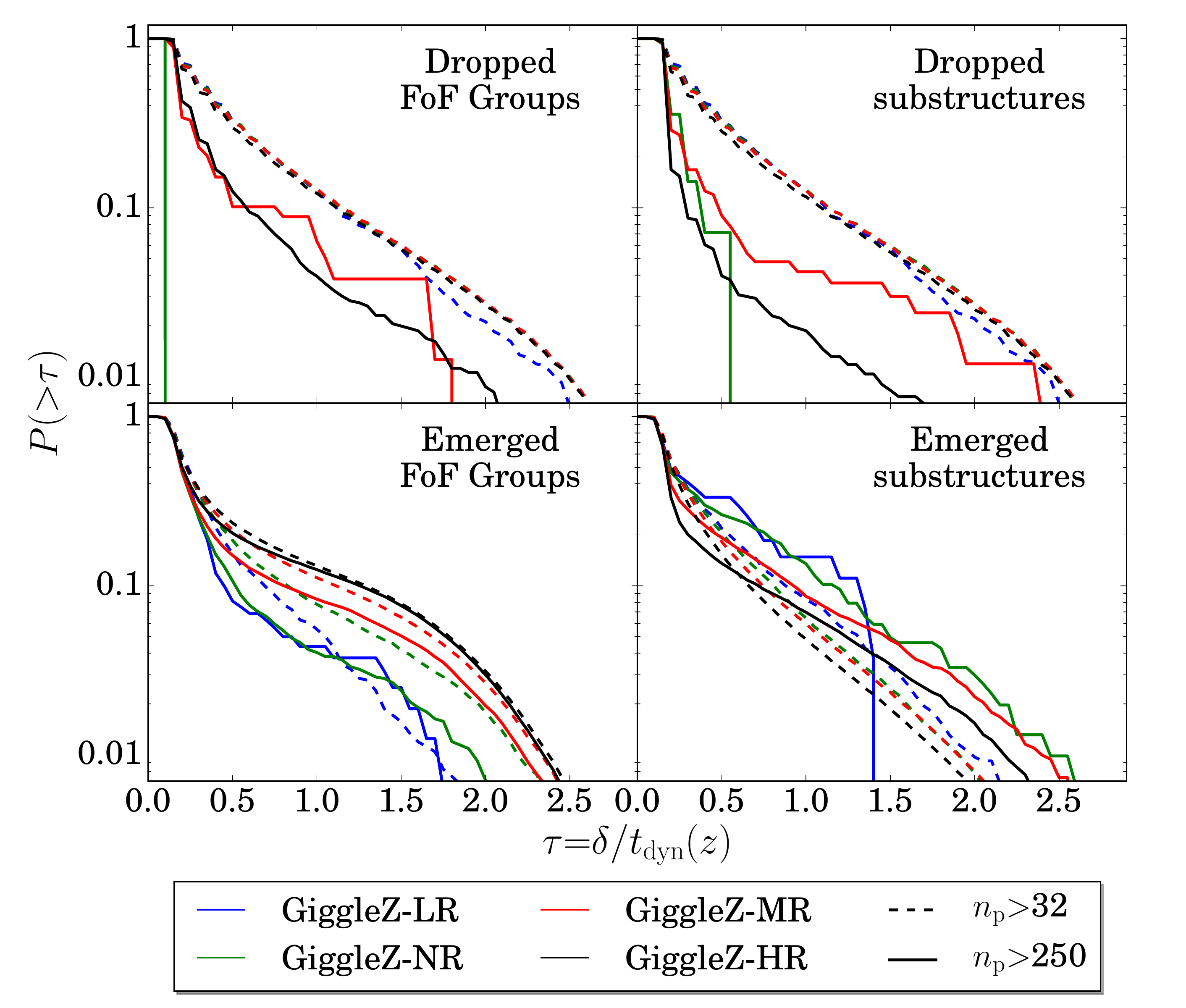}
\caption[Timescales for dropped and emerged haloes]{The distribution of times ($\delta$; expressed in units of a halo's dynamical time \tdyn) between dropped haloes and their descendants [top] and emerged haloes and their progenitors [bottom].  Distributions are expressed as probabilities of exceeding the ordinate value and are shown for both FoF groups [left] and substructures [right].  Each of the 4 \GiggleZcontrol\ simulations are illustrated with varying colours (increasing in mass resolution from \GiggleZLR\ [blue], to \GiggleZNR\ [green], to \GiggleZMR\ [red] to \GiggleZHR\ [black]) for two minimum halo sizes: \nparticle${\geq}250$ [solid lines] and \nparticle${\geq}32$ [dashed lines]. \label{fig-timescales}} 
\end{figure}

In Figure \ref{fig-timescales} we present the distributions of times (expressed in units of dynamical time) between dropped haloes and their descendants and emerged haloes and their progenitors, for both FoF haloes and substructures.  At our highest resolutions,  the timescales over which FoF haloes fall out of \Subfind\ catalogs is significantly longer than for substructures.  For emerged halo recovery times, the distribution is bimodal for both populations with a narrow distribution centred at zero accompanied by an extended population of varying amplitude.  This extended population is peaked and centred at $\tau$\simx{1.5} for FoF haloes and is roughly a power law for substructures.  At our highest mass resolution, $90$\% of emerged FoF haloes are recovered within 1.5 dynamical times while $90$\% of substructures are recovered within only 0.75 dynamical times.

Interestingly, the trend of this extended population with mass resolution for resolved haloes works in opposite directions for FoF groups and substructures: as mass resolution is increased, the population of extended FoF recovery times increases while for substructures it decreases.  This leads to a growing difference in recovery times between the two halo types as mass resolution increases.  For FoF groups, haloes drop-out of the catalog when neighbouring structures become linked.  This occurs with \emph{higher frequency} at higher mass resolution and typically lasts for a roughly \emph{constant interval} similar to a dynamical time (due to the fixed comoving density defining the boundaries of these systems), explaining the trend seen for the extended interval population of FoF haloes.  For substructures, haloes drop-out of our catalogs when their core density (which is resolved to larger values as mass resolution increases) becomes comparable to the background density of the system they are getting lost in.  This occurs with \emph{lower frequency} at higher mass resolution (since the filling factor of the background system decreases with increased density) and lasts for \emph{shorter intervals} (owing to a decreased dynamical time at higher densities), explaining the trend for substructures.

\subsubsection{Branch lengths}

\begin{figure}
\includegraphics[width=84mm]{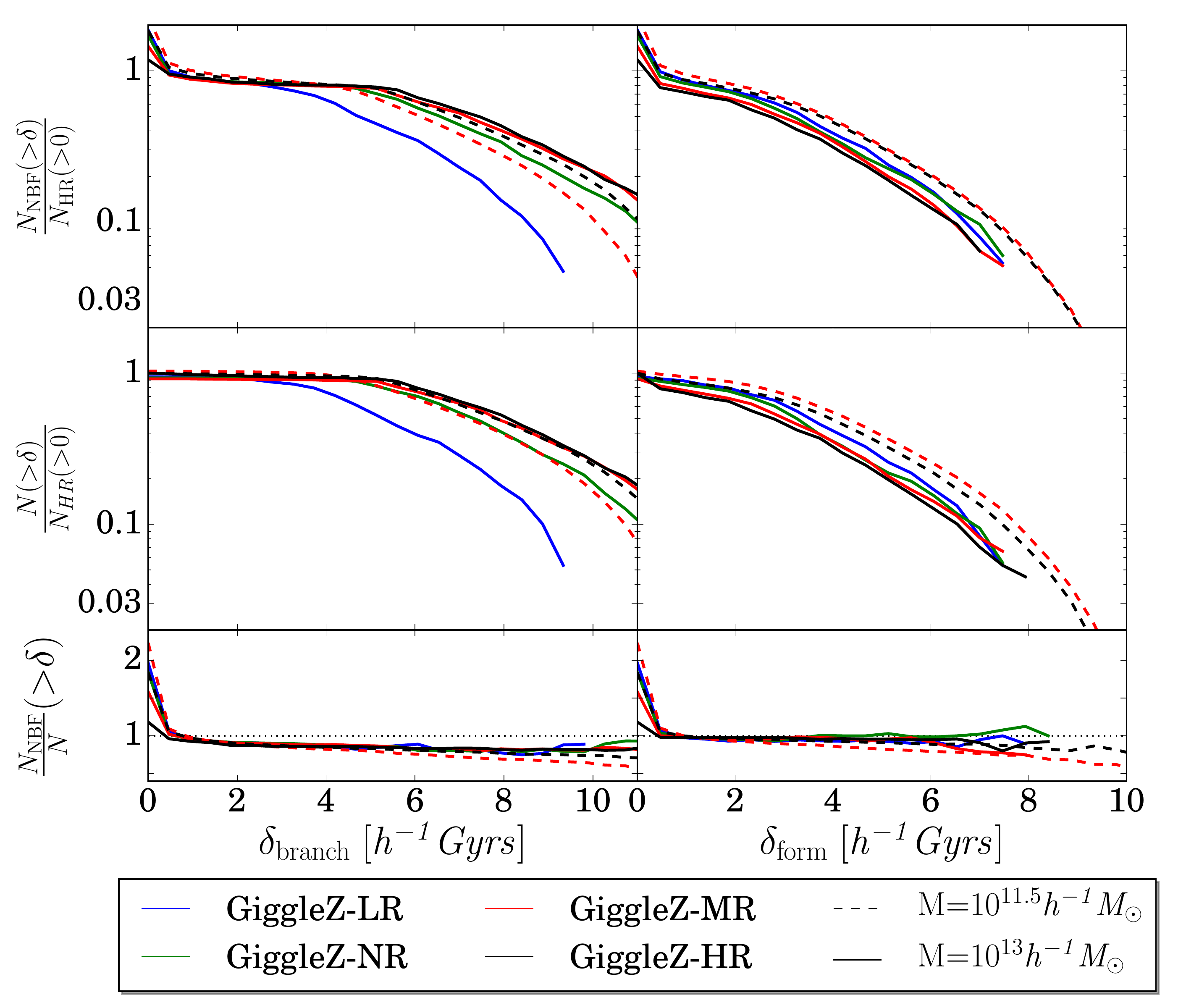}
\caption[Branch ages]{Distribution of FoF halo tree branch ages [left] and half-mass formation times [right] for our nominal tree finding scheme [middle] and the identical case, but with bridged halo fixing turned off [NBF, top]. Each of the 4 \GiggleZcontrol\ simulations are illustrated with varying colours (increasing in mass resolution from \GiggleZLR\ [blue], to \GiggleZNR\ [green], to \GiggleZMR\ [red] to \GiggleZHR\ [black]) for two halo masses ($M{\sim}10^{11.5} {\rm M_\odot}$ [dashed lines] and $M{\geq}10^{13} {\rm M_\odot}$ [solid lines]).  Bottom panels show the fractional difference between the two, highlighting the increased population of very short branches in all NBF cases.\label{fig-branch_ages}} 
\end{figure}

The effect of tree pathologies on FoF merger tree branch lengths has been used as a merger tree diagnostic in several previous studies.  For comparison, we present this here in Figure \ref{fig-branch_ages}.  Results are as expected: turning-off our bridged and emerged halo finding apparatus yields an increase in the number of short branches with lifetimes \lsimx{1} \Gyrs\ and a deficit of branch lengths longer than this.  This is due to merger tree pathologies incorrectly breaking branches of the tree, turning long branches of the trees into two-or-more smaller ones, with a preference for one of these being \lsimx{1} \Gyrs\ in length.  

We also present in this figure a similar plot, but using tree branch formation ages (measured as the time since a halo had a most-massive progenitor with less than half its current number of particles).  We find that this diagnostic performs much better, yielding similar qualitative trends but with much more consistent behaviour across all mass resolutions

Since we present cumulative counts in this figure, a declining linear trend in the residual (bottom panel) with large values of $\delta_{\rm branch}$ is expected if -- in the case where pathology corrections are not applied -- occurrences of breaking a branch are randomly distributed amongst the branches of the trees.  This is consistent with an accumulation of events over time for haloes of all sizes and reenforces the fact that merger tree pathologies are a serious problem for large and well resolved (or old) haloes, as well as small and less resolved ones.

\subsection{Merger rates}\label{sec-merger_rates}

\begin{figure*}
\begin{minipage}{170mm}
\begin{center}
\includegraphics[width=170mm]{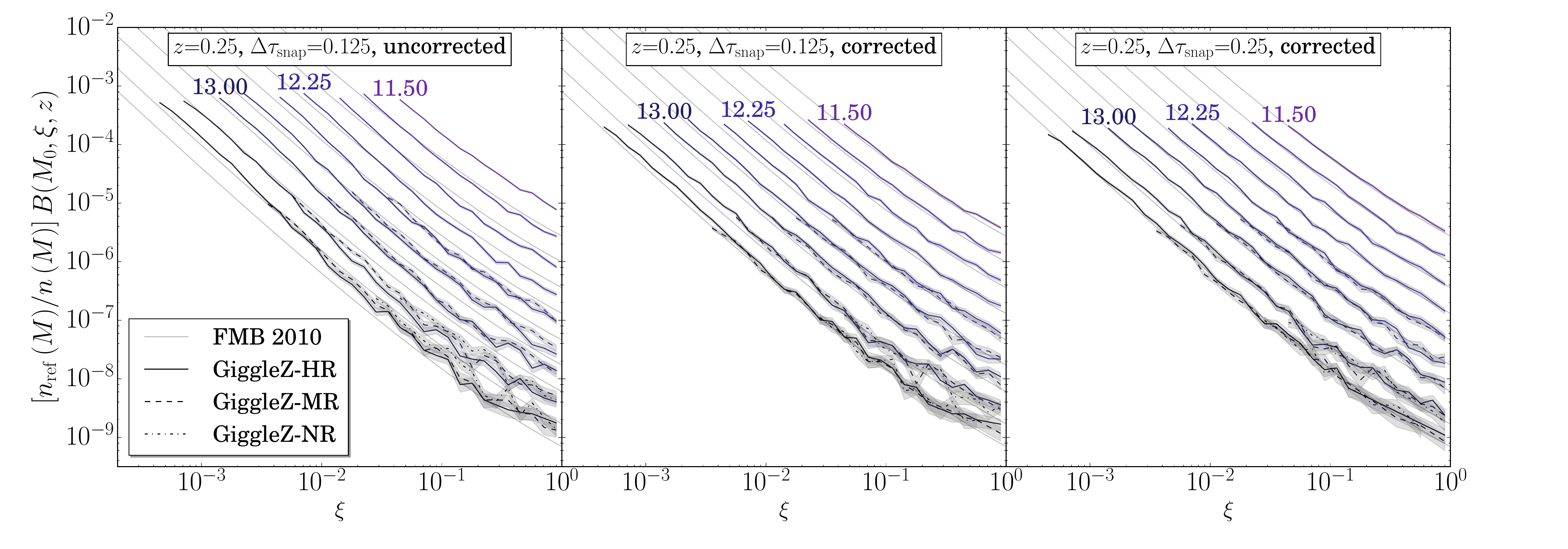}
\caption{Friends-of-friends halo merger rates as a function of merger ratio averaged over a $\delta z{=}0.5$ interval centred on \zxeqx{}{0.25} for the \GiggleZcontrol\ simulations.  Three cases are shown: two where our nominal snapshot cadence is used with [left] no correction for merger tree pathologies or [middle] full corrections for merger tree pathologies.  These are compared to [right] the case where a snapshot cadence comparable to that of the \Millennium\ simulation is used, with full pathology corrections.  The analytic fits to the \Millennium\ simulation presented by \citet[][FMB10]{Fakhouri:2010p1134} are shown with grey lines.  Poisson uncertainties are shown as shaded grey regions.  Remnant halo masses ($\log M[M_{\odot}]$) are indicated for each curve.  As in FMB10, only mergers with remnant particle counts \nparticle${>}500$ and secondary particle counts \nparticle${>}40$ are displayed.  All curves are normalised to the halo abundance of the \GiggleZHR\ simulation ($n_{\rm ref}$). \label{fig-rates_millennium}}
\end{center}
\end{minipage}
\end{figure*}

Lastly, we turn now to an analysis of merger rates expressed in terms of redshift, merger remnant mass\footnote{Merger rates are typically expressed in terms of remnant mass rather than (for example) primary mass, to facilitate comparisons to extended Press-Schechter theory wherein merger rates are most naturally expressed in these terms.} and mass ratio.  We follow the approach first established by FM08 and subsequently utilised by \citet[FMB10]{Fakhouri:2010p1134} and \citet{Genel:2010p2670} in their analyses of the \Millennium\ simulations.  For every remnant halo in a set of merger trees with $N_{\rm proj}{>}1$ progenitors, we sort its progenitors by declining mass (denoted as \rmsub{M}{i} such that \rmsub{M}{0}{$>$}\rmsub{M}{i}{$>$}\rmsub{M}{N_{\rm proj}-1}).  As shown by FM08, progenitors $i{>}0$ can be treated as a sequence of independent binary mergers involving a primary system of mass \rmsub{M}{0}, a secondary system of mass \rmsub{M}{i}, a remnant system of mass \rmsub{M}{}{$=$}\rmsub{M}{0}{$+$}\rmsub{M}{i} and a merger ratio of $\xi_{\rm i}{=}M_{\rm i}/M_{\rm 0}$.  FM08 established that the mean merger rate per halo, when expressed as a function of merger ratio, remnant mass and redshift, has a nearly universal dependance on $\xi$ with a normalisation that is insensitive to remnant mass and redshift:
\begin{equation}\label{eqn-FM10_eqn1}
\frac{B(M,\xi,z)}{n(M,z)} = A\left(\frac{M}{10^{12}M_\odot}\right)^{\alpha}\left(1+z\right)^{\eta}\Xi(\xi)
\end{equation}

\noindent The factor describing the merger ratio dependance is given by
\begin{equation}\label{eqn-FM10_eqn1_xi_dependance}
\Xi(\xi) = \xi^{\beta} \exp{\left[\left(\frac{\xi}{\tilde{\xi}}\right)^{\gamma}\right]}
\end{equation}

\noindent with updated best-fit values presented by FMB10 given in Table \ref{table-merger_rate_fit_parameters}.

In Figure \ref{fig-rates_millennium} we compare FoF halo merger rates, extracted following this approach from the \GiggleZ\ simulations, to the analytic fit to the \Millennium\ simulations presented in FMB10.  This is done for three cases: two where we use our nominal snapshot cadence (\Deltatausnap\eqx{0.125}) and make no attempt to correct for tree pathologies (left) or apply all tree pathology corrections (middle), and another where we use a snapshot cadence approximately the same as that used for the \Millennium\ simulation (\Deltatausnap\eqx{0.25}) with full corrections for tree pathologies (right).  We note several things from this figure.  First, we find excellent agreement with the results of FMB10 when we apply our pathology corrections, validating their \quotes{stitch} approach to correcting tree pathologies.  However, the snapshot cadence of the \Millennium\ simulation (which does not meet our snapshot cadence convergence criterion) yields a slight downward bias in merger rates.  When we use a snapshot cadence comparable to that of \Millennium, our measurements are brought into even better agreement with the FMB10 fits.  Second, the substantial bias towards higher merger rates found in Section \ref{sec-convergence_mass_res} when tree pathologies are not corrected for is clearly illustrated in the left panel.  Again, it is important to note that this bias is present for all remnant mass bins and merger ratios, reenforcing the fact that tree pathologies are an important problem even for the most well-resolved haloes of a simulation.  Lastly, we find excellent agreement between the the merger rates obtained from the \GiggleZNR, \GiggleZMR\ and \GiggleZHR\ simulations (\GiggleZLR\ does not have enough resolution to populate this plot for these halo size cuts) illustrating excellent agreement across a factor of 64 in mass resolution.

\begin{figure*}
\begin{minipage}{170mm}
\begin{center}
\includegraphics[width=170mm]{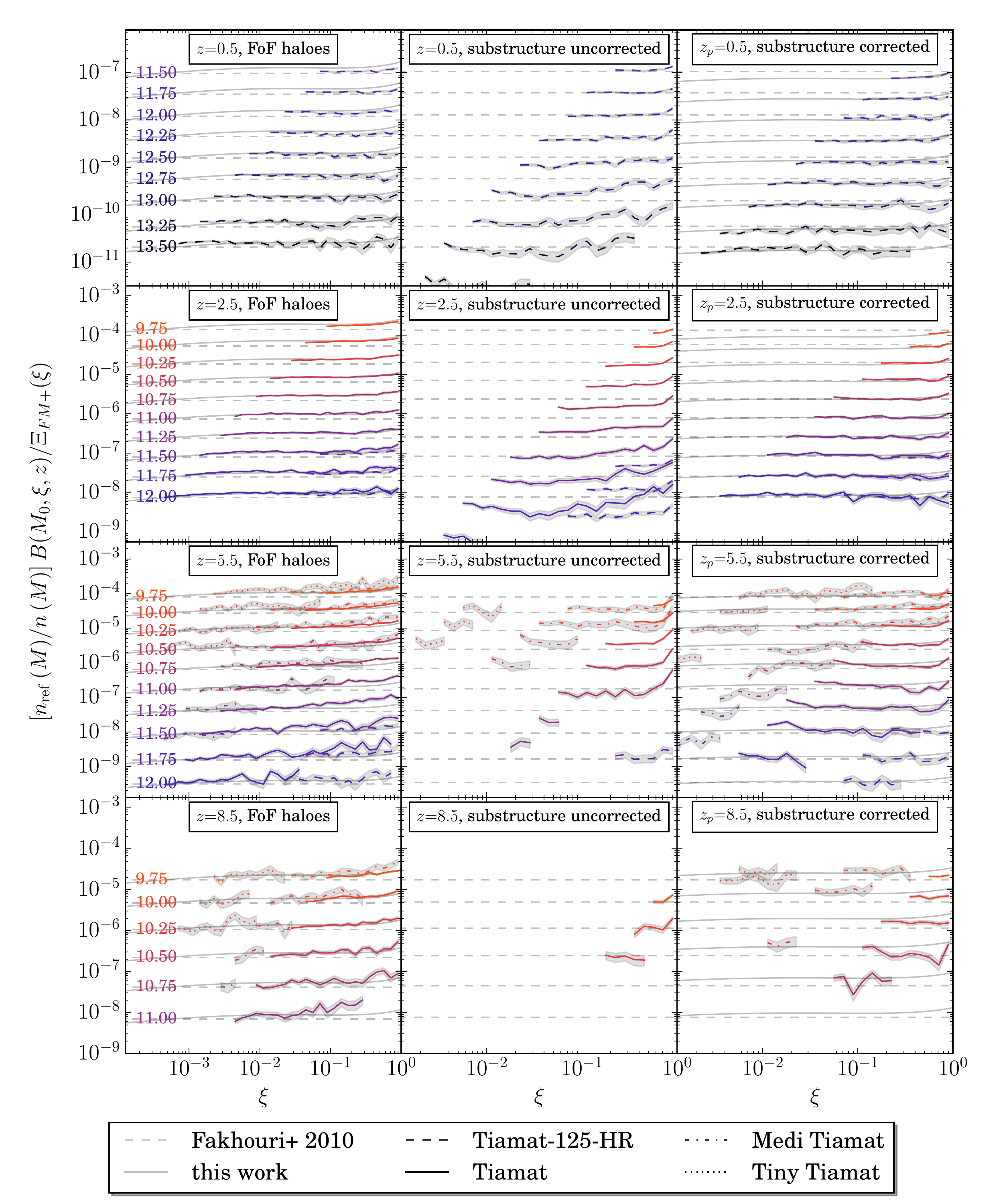}
\caption{Merger rates as a function of merger ratio for [left panels] FoF haloes and [middle/right panels] substructures at several redshifts ($\delta z{=}1$, centred on the given value) covering the range adequately resolved by our simulations. Uncorrected substructure rates [middle] are those extracted directly from the trees while corrected substructure rates [right] include a correction calculated from the satellites present in the last snapshot of the simulation. The merger ratio dependance, as determined by \citet{Fakhouri:2010p1134} ($\Xi_{\rm FM+}$) has been divided-out to emphasise the slight differences between this work and theirs.  The analytic fits to the \Millennium\ simulation presented by \citet{Fakhouri:2010p1134} are shown with dashed grey lines.  The analytic fits found in this work are shown with grey solid lines.  Poisson uncertainties are shown as shaded grey regions.  Remnant halo masses ($\log M[M_{\odot}]$) are indicated for each curve.  Only mergers with remnant particle counts \nparticle${>}1000$ and secondary particle counts \nparticle${>}75$ or \nparticle${>}100$ are displayed for FoF or substructures respectively.  For cases where multiple simulations resolve the same ($z$,$\log M$)-pair, curves are corrected to the halo abundance of the highest-resolution simulation ($n_{\rm ref}$).\label{fig-rates_multiz}}
\end{center}
\end{minipage}
\end{figure*}

We have extracted FoF and substructure halo merger rates from the \Tiamat\ simulations at redshifts $z{=}{0.5}{\pm}{0.5}$, ${2.5}{\pm}{0.5}$, ${5.5}{\pm}{0.5}$ and ${8.5}{\pm}{0.5}$ (covering nearly the full range of observed galactic history; we do not use $z{>}8.5$ since our snapshot cadence rises above our convergence criterion at that point) for systems with minimum remnant particle counts of \nparticle\gtrx{1000} and secondary particle counts of \nparticle\gtrx{75} and \nparticle\gtrx{100} for FoF and substructure haloes respectively.  In Figure \ref{fig-rates_multiz} we present the results.  To add clarity, we have divided-out the merger ratio dependence ($\Xi$; Equation \ref{eqn-FM10_eqn1_xi_dependance}) using the parameters of the FMB10 fits.  This cleanly expresses the mass and redshift dependence of the plotted merger rates while emphasising slight deviations in the mass ratio dependence of our merger rates from that of FMB10.

\begin{table*}
\begin{minipage}{170mm}
\begin{center}
\begin{tabular}{ccccccccc}
\multicolumn{9}{c}{FoF haloes}	\\
\cmidrule(lr){1-9}
							&
\multicolumn{2}{c}{Best Fit Values}	& 
\multicolumn{6}{c}{Covariance Matrix}		\\
\cmidrule(lr){2-3}                  
\cmidrule(lr){4-9}
Parameter		&
FMB10		& 
This work   	&
A 			&
$\tilde{\xi}$	&
$\alpha$ 		&
$\beta$ 		&
$\gamma$	&
$\eta$   \\
\hline
A			&	0.0104	&	0.107	&	\scinote{3.84}{-6}  & \scinote{2.37}{-5}  & \scinote{1.52}{-7}  & \scinote{5.51}{-6}  & \scinote{1.98}{-5}	& \scinote{-1.33}{-7}	\\
$\tilde{\xi}$	&	0.00972	&	0.483	&	\scinote{2.37}{-5}  & \scinote{1.52}{-4}  & \scinote{5.90}{-7}  & \scinote{3.45}{-5}  & \scinote{1.27}{-4}	& \scinote{2.74}{-6}	\\
$\alpha$		&	0.133	&	0.122	&	\scinote{1.52}{-7}  & \scinote{5.90}{-7}  & \scinote{7.84}{-7}  & \scinote{5.85}{-7}  & \scinote{-1.51}{-7}& \scinote{2.08}{-6}	\\
$\beta$		&	-1.995	&	-1.751	&	\scinote{5.51}{-6}  & \scinote{3.45}{-5}  & \scinote{5.85}{-7}  & \scinote{8.48}{-6}  & \scinote{2.78}{-5}	& \scinote{1.82}{-6}	\\
$\gamma$	&	0.263	&	0.670	&	\scinote{1.98}{-5}  & \scinote{1.27}{-4}  & \scinote{-1.51}{-7} & \scinote{2.78}{-5}  & \scinote{1.18}{-4}	& \scinote{7.28}{-7}	\\
$\eta$		&	0.0993	&	0.069	&	\scinote{-1.33}{-7} & \scinote{2.74}{-6}  & \scinote{2.08}{-6}  & \scinote{1.82}{-6}  & \scinote{7.28}{-7}	& \scinote{1.07}{-5}	\\
\\
\multicolumn{9}{c}{Substructure}	\\
\cmidrule(lr){1-9}
A			&	-	&	0.039	&	\scinote{6.50}{-8}  & -  & \scinote{-9.94}{-8} & -  & - & \scinote{-1.34}{-6}	\\
$\alpha$		&	-	&	0.254	&	\scinote{-9.94}{-8} & -  & \scinote{1.86}{-6}  & -  & - & \scinote{5.81}{-6}	\\
$\eta$		&	-	&	0.872	&	\scinote{-1.34}{-6} & -  & \scinote{5.81}{-6}  & -  & - & \scinote{3.93}{-5}	\\
\hline
\end{tabular}
\caption{Best-fit parameters for the merger rate parameterisation given by Equation \ref{eqn-FM10_eqn1} from the fit to the \Millennium\ simulations presented by \citet[FMB10]{Fakhouri:2010p1134} and to the \Tiamat\ simulations (this work).  Note that little data with $\xi{<}10^{-3}$ was used for the fit in this work, so our parameterisation should not be applied to merger ratios less than this.  The covariance matrix for the fit presented in this work is also given.\label{table-merger_rate_fit_parameters}}
\end{center}
\end{minipage}
\end{table*}

From the left column of this figure we see once again that our analysis of FoF merger rates is broadly in very good agreement with the results presented by FMB10.  Interesting differences are found, however.  While the amplitude of minor-merger rates is very similar, the merger ratio dependance is slightly flatter than that of FMB10 with more major mergers obtained from our analysis.  Very good agreement is seen between simulations of varying resolutions and box sizes, although anecdotally, there seems to be a systematic trend for the abundance of the most massive major mergers of each simulation to be biased high.  Presumably, this is a product of our simulations' inability to properly capture tidal effects on the scales of their periodic boxes, resulting in incorrect orbits and/or biased mass accretion histories for these systems.

In the middle column (labelled \quotes{substructure uncorrected}), we present substructure merger rates as obtained from counting all the non-fragmented progenitors in the trees.  At lower masses we see that these rates roughly track those of the FoF haloes and show reasonable agreement across various mass resolutions.  At higher masses, rates drop precipitously relative to the FoF rates and agreement between resolutions is lost.  These trends can be traced to the fact that substructures (even central substructures) take a significant period of time to merge subsequent to the accretion of their host FoF haloes onto other systems (\cf\ the difference between the magenta and orange histograms in Figure \ref{fig-M_tracks}).  Towards the end of a simulation, this delay can extend past the last snapshot resulting in accreted substructures not being encoded as mergers in the trees.  Furthermore, since the delay between a substructure's accretion and its final merger with its host is significantly resolution-dependant, this results in significant resolution dependance in substructure merger rates when extracted from the progenitor lists of the trees.

Since we are interested here in studying the convergence properties of our substructure merger trees, we have corrected our substructure merger rates to account for this effect.  We do so by defining substructure mergers as events occurring at the point when they reach their peak halo size.  With this ansatz, a robust measure of substructure merger rates can be obtained by assuming all substructures present in the last simulation snapshot will merge with their central halo\footnote{Satellite-satellite mergers are rare, making this a reasonable and accurate assumption.}, adding these to the mergers encoded by the progenitors of our trees, and binning them in redshift by $z_p$.  We present the results in the right column of Figure \ref{fig-rates_multiz} (labelled \quotes{substructure corrected}) and see now that the rates behave much more like the FoF merger rates and exhibit considerably more agreement across resolutions.

Using the Monte-Carlo Markov Chain code first presented in \cite{Poole:2013p1849}, we have performed fits to both our FoF and corrected substructure merger rates and present the results in Table \ref{table-merger_rate_fit_parameters}.  No complicated degeneracies between the parameters in the resulting posterior distribution function are found.  Note that very little data with $\xi{<}10^{-3}$ was used for this fit, so our parameterisation should not be applied to merger ratios less than this.  For substructure rates, there is little suggestion that the merger rate dependence on $\xi$ differs significantly between FoF and substructure haloes, so we have fixed the parameters that Equation \ref{eqn-FM10_eqn1_xi_dependance} depends on to their FoF values for the corrected substructure fits.  

By fixing the mass-ratio dependance of the substructure merger rate parameterisation, we simplify the comparison of the parameters expressing the mass and redshift dependance ($\alpha$ and $\eta$ respectively) between the two populations.  These parameters show what can be seen from Figure \ref{fig-rates_multiz}:  there is a significantly stronger dependence on both for the substructure merger rates than for the FoF rates.  While we leave a detailed account of this to future study, initial investigations indicate that this is a product of variations in substructure fractions and in the channels through which substructure can evolve \citep[see][for an account]{vandenBosch:2016p2681}.

We find that the FoF merger rates obtained in this work reduce the discrepancies found in previous works from extended Press-Schechter (EPS) estimates.  In Figure \ref{fig-rates_EPS_compare} we compare the analytic fits of FM08, FMB10 and that found in this work to the \quotes{Option B} form of the formula for EPS merger rates given in FM08 (equation 10 of that work).  We find that the substantial excess of EPS major mergers reported by FM08 (and sustained by the updated fits of FMB10) are reduced.  Similar excesses of minor mergers in the simulations to those seen in previous studies are found.  The trend of the discrepancy with merger ratio is much flatter overall than that of either FM08 or FMB10.

Lastly, we note at this point that a systematic problem arrises at the end of a simulation where the interval available for generating back matches declines and shrinks to zero.  For \gbpTrees\ and tree-building methodologies like it, this compromises our ability to correct for pathologies and systematically biases merger rates.  In Figure \ref{fig-rates_compare_z0pt0} we illustrate the magnitude of this effect by comparing the instantaneous merger rates in the \TiamatHR\ simulation for secondary haloes with halo particle sizes with \nparticle${\ge}50$, ${\ge}500$ and ${\ge}5000$ for two sets of merger trees: one run to the final snapshot at \zxeqx{last}{0} (illustrated in black) and one stopped exactly one scan interval before this at \zxeqx{last}{0.21} (illustrated in red).  Towards the end of each set of trees, a distinct rise in merger rates is seen, counter to the mildly declining trend.  The magnitude of the effect can be seen in the bottom panel where the difference between the two sets of trees is presented.  For haloes with \nparticle${\sim}100$ particles, the instantaneous merger rates are discrepant at a level as high as 100\%.  For haloes with \nparticle${\ge}5000$ particles, the effect is minimal.  We have repeated this exercise at \zxeqx{last}{5} with the \Tiamat\ simulation (not shown) and obtained  very similar results.  Importantly, all merger rates obtained near the ends of our simulations which we have so-far reported are averaged over redshift intervals sufficiently large to render this effect unimportant.  However, given the lofty ambitions in precision for future cosmology studies discussed in the introduction, this is an effect that may need to be considered.  The most straight-forward solution would be to run simulations for a full scan interval past \zxeqx{}{0} (\ie\ to an expansion factor of $a_{\rm last}{\sim}1.2$ or a redshift $z_{\rm last}{\sim}{-}0.2$).

%% file: discussion.tex
We have demonstrated that -- with careful treatment -- accurate and robust halo and substructure tracking is possible for catalogs generated from configuration-space halo finders.  While it is true that the use of phase-space halo finders will likely yield improved halo tracking results, it is still unclear at this point precisely how significant this improvement is likely to be.  For applications in galaxy formation studies, it is quite possible that systematic differences in substructure masses, spins, \etc\ are far more relevant.  

This is almost certainly the case at high redshifts.  In \citet{Poole:2016p2664} it has been demonstrated that the top-level substructure hierarchy of FoF haloes is substantially and ubiquitously different between \Subfind\ and \ROCKSTAR, owing to the dynamically long periods of time that massive substructures retain their identities in phase space following mergers, and the dynamically short intervals between major merger events \citep[although, the phase-space halo finder \Velociraptor\ appears to perform much better in this regard; see][for details]{Elahi:2011p2683,Behroozi:2015p2660}.  Our initial testing at high-redshift with the semi-analytic model \Meraxes\ \citep{Mutch:2016p2671} indicates that such differences will require significant changes to model parameters and/or prescriptions for baryonic cooling, star formation, disk growth, \etc. to permit accurate models using both catalogs.  While just one model, \Meraxes\ is similar to many in the literature suggesting that this may be a broad issue.

Ultimately, careful direct comparisons of hydrodynamic simulations to semi-analytic methods \citep[see][or Qin \etal\ in prep., for an example]{Stevens:2016p2668} applied to halo catalogs constructed with a variety of halo finding methods is likely to produce the clearest answers regarding the best path towards the most robust, accurate and physically insightful numerical approaches and physical prescriptions for semi-analytic models.  Halo tracking -- when carefully done -- is not likely to be a substantial systematic in this pursuit.

%% file: summary.tex
The process of halo matching across the snapshots of cosmological simulations is of fundamental importance to the task of constructing merger trees from halo catalogs.  We have motivated the need to ensure that this process is optimised so that pointers between haloes robustly follow core material, to avoid problems introduced by situations which can confuse naive matching schemes.  These situations include central-satellite switching, the ejection of halo cores during dynamical disturbances and spurious matches to small haloes which pass directly through the centres of larger haloes.  We present the details of a procedure using cuts in a 2-dimensional space constructed from pseudo-radial moments of radially-sorted halo particle lists which effectively ensures this.  This process is controlled by one parameter, denoted \Deltafgood.

With the efficacy of our approach to this fundamental mechanic of tree building firmly established, we have then presented a phenomenological approach to classifying pathologies in merger trees introduced by inadequacies in the halo finding process.  Several classes of pathological haloes are defined as a result: strayed (\ie\ descendant-less) haloes, emerged haloes (those found to enter another halo, become lost, and re-emerge later) and fragmented haloes (those found to exit another halo without being assigned a progenitor).  These classifications are based on the network of matches that can be constructed both to-and-from haloes over a range of snapshots (\ie\ the \quotes{search} interval, denoted \Deltatausearch\ with units of dynamical time) extending both forward-and-backward in time.  We present the \gbpTrees\ algorithm which is designed to build robust merger trees capable of treating these cases.

\begin{figure}
\begin{center}
\includegraphics[width=75mm]{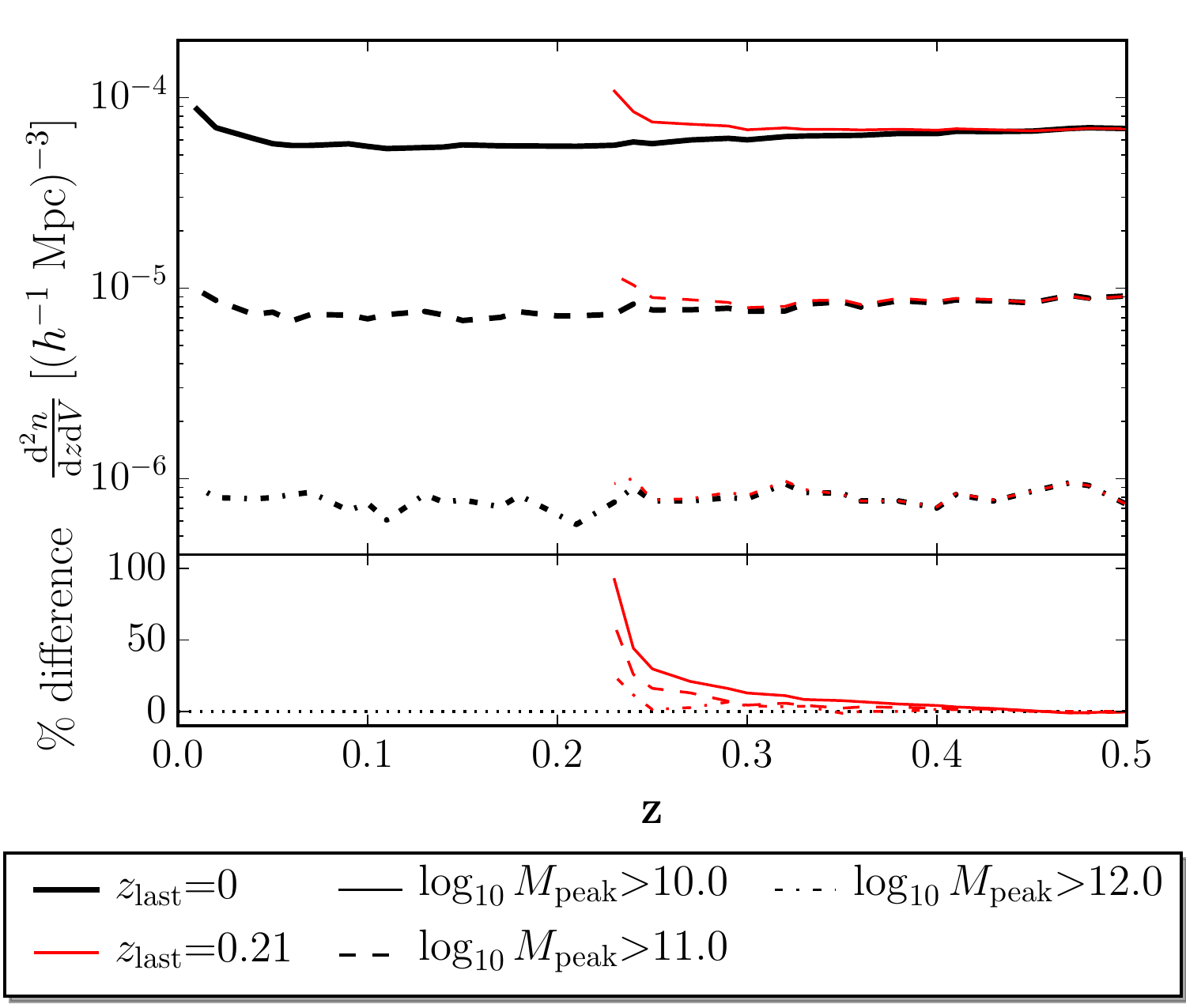}
\caption{A comparison of merger rates (for secondary haloes larger than three peak mass cuts corresponding to particle counts ${\ge}50$, ${\ge}500$ and ${\ge}5000$ from top-to-bottom) in the \TiamatHR\ simulation for two cases: \zxeqx{last}{0} and $0.21$ (exactly one fiducial scan interval prior to \zxeqx{}{0}).  Both cases show an increase in merger rates close to $z_{\rm last}$ with the bottom panel illustrating the size of the effect for the $z_{\rm last}{=}0.21$ case.  This demonstrates the effect the absence of back-matches towards the simulation's end has on merger rates, suggesting that for optimal results at the current epoch, simulations should be run a full scan interval past \zxeqx{}{0}.\label{fig-rates_compare_z0pt0}}
\end{center}
\end{figure}

We then consider the convergence behaviour of total merger counts and tree pathology populations for this algorithm under variations in \Deltafgood, \Deltatausearch\ and \Deltatausnap\ (where \Deltatausnap\ represents the interval between snapshots in units of dynamical time) when applied to halo catalogs constructed with \Subfind:

\begin{itemize}
\item Firstly, we establish the strictest value of \Deltafgood\ which yields convergence for these populations.  We then find that significantly longer search intervals (\Deltatausearch\eqx{2}) are required for converged FoF trees while much shorter search intervals (\Deltatausearch\eqx{1}) are adequate for converged subhalo trees.  

\item Next, the ideal snapshot cadence is found to be \Deltatausnap\eqx{0.128}.  This results in FoF and subhalo merger counts which are converged at a level of \lsimx{5\%} over nearly the full mass range where we have adequate statistics.  However, fragmented haloes, as we presently define them, illustrate poor convergence as \Deltatausnap\ is extended to values \lsimx{0.128}.  This snapshot cadence is conveniently similar to that found necessary by \citet{Benson:2012p2682} for the convergence of galaxy masses in semi-analytic models to a level of 5\%. 

\item Lastly, we examine the convergence of these populations with mass resolution under the assumption of these fiducial values (\Deltafgood,\Deltatausearch,\Deltatausnap)${=}(-0.2,2,0.128)$.  We find that total merger counts agree across a factor of 512 in mass resolution for FoF haloes at a level of \lsimx{5}\% for haloes with \nparticle\gsimx{75} particles.  Subhalo total merger counts exhibit poor convergence at low mass resolutions but converge at a level of \simx{10}\% for haloes with \nparticle\gsimx{100} particles at mass resolutions of $m_{\rm p}{\lsim}{10}^{9}$ \Msolhunit.  Strayed haloes are found to be restricted to haloes with \nparticle\lsimx{100}. Fragmented haloes exhibit slow convergence with mass resolution, reaching levels of agreement of only \simx{50}\% at highest mass resolutions, although such cases are rare.  Fragmented FoF haloes converge in mass suggesting a physical origin while fragmented substructures exhibit their convergence in particle count, suggesting a numerical origin. 
\end{itemize}

When characterising the masses of subhaloes, we argue that the peak mass reached by the subhalo is the best metric (for our \Subfind\ catalogs at least), since this minimises systematic biases with mass resolution and snapshot cadence for instantaneous measures.  Central-satellite switching events introduce a serious bias in this measure however and we present a \quotes{dominant} halo mechanism to deal with this.  When a new progenitor-less FoF halo is established in the trees, the central subhalo at that point is designated as its dominant subhalo.  This designation is propagated forward along the dominant halo's main progenitor line.  When mergers between FoF haloes occur, the most massive dominant halo is propagated forward as the remnant's dominant halo.  Only situations where a central halo is a dominant halo are used for peak halo size calculations, eliminating the unwanted effects of large transient increases in the sizes of subhaloes due to central-satellite switching events.

\begin{figure}
\begin{center}
\includegraphics[width=75mm]{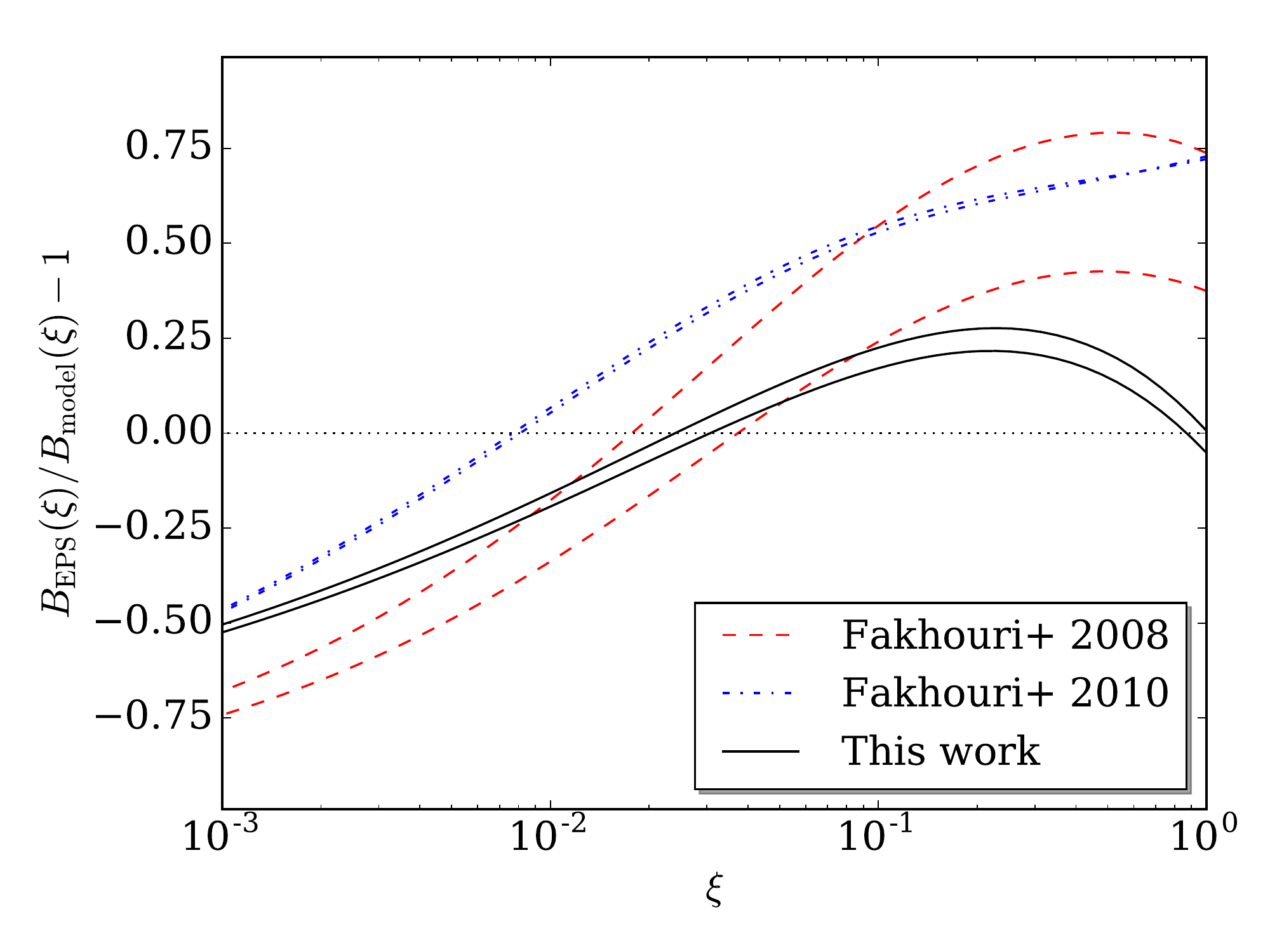}
\caption{A comparison of \zxeqx{}{0} merger rates given by the analytic fits to simulation results obtained in \citet{Fakhouri:2008}, \citet{Fakhouri:2010p1134} and in this work, to the expression for extended Press-Schechter (EPS) merger rates (option B) derived in \citet{Fakhouri:2008}.  In each case, top lines are the relations for \TTTP{14}\Msol\ haloes and bottom lines are the relations for \TTTP{12}\Msol\ haloes.\label{fig-rates_EPS_compare}}
\end{center}
\end{figure}

Taking these issues into account, we provide fits to the merger rates of both FoF and substructure haloes across nearly all of observed cosmic time.  Results for FoF haloes are very similar to those reported by FMB10, with the exception that our analysis yields a slightly higher incidence of major mergers.  This is found to decrease the discrepancies reported in previous studies between the merger rates computed from EPS analyses and simulations.  When substructure mergers are defined as occurring at the point where they reach peak halo size (and with a correction for extant satellites in the final simulation snapshot), the mass-ratio dependence of their rates are found to be very similar to those of FoF haloes.  Significantly stronger dependencies of their merger rate normalisation on mass and redshift are found however.

We emphasise again that all these results are intended only for the interpretation of merger trees built from halo catalogs extracted specifically with the \Subfind\ halo finder (with potential but likely varying degrees of applicability to other configuration-space halo finders).  Phase-space halo finders will certainly be less affected by these pathologies, but not immune.  Since the \gbpTrees\ algorithm takes only the particle IDs of a halo catalog (radially sorted in some way) as input, our algorithm should be equally applicable to catalogs extracted with any halo finder.  

For merger tree construction approaches exploying particle matching for building tree pointers, the need for matches which traverse both backwards and forwards in time to correct for tree pathologies suggests that simulations should be run past the last epoch needed for intended analyses.  For studies at redshift \zxeqx{}{0} for example, we argue that simulations should be run to expansion factors of $a_{\rm last}{\sim}1.2$, or a redshift $z_{\rm last}{\sim}{-}0.2$.

Our explorations at high redshift indicate that phase-space halo finders can split the substructure hierarchy of collapsing regions very differently, suggesting significantly different late-stage halo properties and accretion histories.  Since the underlying premise of semi-analytic galaxy formation models is that halo properties extracted from collisionless cosmological simulations can be accurately mapped onto the baryonic assembly of real galaxies through sensible and insightful physical prescriptions, such differences are likely to be important in our pursuit of more robust and accurate galaxy formation models in the coming era of massive galaxy surveys.  It is unclear at this point which approach best facilitates this and much more study on this issue is needed.  However, the issue of halo tracking is ultimately not likely to be significant in the course of deciding this matter.

%% file: thanks.tex
This research was supported by the Victorian Life Sciences Computation Initiative (VLSCI), grant ref. UOM0005, on its Peak Computing Facility hosted at the University of Melbourne, an initiative of the Victorian Government, Australia. Part of this work was performed on the gSTAR national facility at Swinburne University of Technology.  gSTAR is funded by Swinburne and the Australian Governments Education Investment Fund. This research program is funded by the Australian Research Council through the ARC Laureate Fellowship FL110100072 awarded to JSBW.